# In-flight performance of the Canadian Astro-H Metrology System

**Luigi C. Gallo,**[a,*] **Alexander Koujelev,**[b] **Stéphane Gagnon,**[c] **Timothy Elgin,**[c] **Martin Guibert,**[c] **Ryo Iizuka,**[d] **Manabu Ishida,**[d] **Kosei Ishimura,**[d] **Naoko Iwata,**[d] **Taro Kawano,**[d] **Casey Lambert,**[a] **Franco Moroso,**[b] **Shiro Ueno,**[d] **Takayuki Yuasa**[e]

[a] Saint Mary's University, Department of Astronomy & Physics, 923 Robie Street, Halifax, Canada, B3H 3C3
[b] Canadian Space Agency, 6767 route de l'Aeroport rd., St-Hubert, QC, Canada, J3Y 8Y9
[c] Neptec Design Group, 302 Legget Dr., Kanata, Canada, K2K 1Y5
[d] Japan Aerospace Exploration Agency, Institute of Space and Astronautical Science, 3-1-1 Yoshino-dai, Chuo-ku, Sagamihara, Kanagawa, Japan, 252-5210
[e] Nishina Center for Accelerator-based Science, RIKEN, 2-1 Hirosawa, Wako, Saitama, Japan, 351-0198

**Abstract**. The Canadian Astro-H Metrology System (CAMS) on the Hitomi X-ray satellite is a laser alignment system that measures the lateral displacement (X/Y) of the extensible optical bench (EOB) along the optical axis of the hard X-ray telescopes (HXTs). The CAMS consists of two identical units that together can be used to discern translation and rotation of the deployable element along the axis. This paper presents the results of in-flight usage of the CAMS during deployment of the EOB and during two observations (Crab and G21.5-0.9) with the HXTs. The CAMS was extremely important during the deployment operation by providing real-time positioning information of the EOB with micrometer scale resolution. In this work, we show how the CAMS improves data quality coming from the hard X-ray imagers. Moreover, we demonstrate that a metrology system is even more important as the angular resolution of the telescope increases. Such a metrology system will be an indispensable tool for future high resolution X-ray imaging missions.

**Keywords**: Hitomi (Astro-H), metrology system, telescope alignment, optics, lasers, CAMS.

*Luigi Gallo, E-mail: lgallo@ap.smu.ca

## 1 Introduction

The Hitomi X-ray mission[1] (formerly called Astro-H) was launched on February 17, 2016 from Tanegashima Space Center by the Japan Aerospace Exploration Agency (JAXA). The international mission was carried out in collaboration with the National Aeronautics and Space Administration (NASA), the European Space Agency (ESA) and the Canadian Space Agency (CSA). The mission included several instruments (four telescopes and five detectors) geared to explore the high-energy Universe between 0.3 – 600 keV[1].

Two of the four telescopes were Hard X-ray Telescopes[2] (HXTs), each focused photons with energies between 5 – 80 keV to identical imaging detectors called the Hard X-ray Imager[3] (HXI) located 12-meters away. To achieve the long focal length, without compromising the compact launch package, an extensible optical bench (EOB) was used. The HXI detectors were placed at the end of the 6-meter EOB. Once extended, the EOB would be subject to distortions primarily from thermal fluctuations in low-Earth orbit (LEO) that would degrade HXI image quality. With the desired objective of providing good quality images, the effect of the distortion needs to be corrected.

The CSA contribution to the mission was the Canadian Astro-H Metrology System (CAMS). The CAMS was a laser alignment system that measured lateral (X and Y) displacement along the optical bench between the HXTs and HXIs. Two identical CAMS units were installed and used in conjunction to provide the capability to measure lateral translation and rotation in the optical bench. These measurements were used to correct and enhance the images obtained with the HXIs.

Much of the efforts leading up to the CAMS flight system appeared in earlier work. The original concept was introduced in Ref. 4 while hardware testing and initial calibration appeared in Ref. 5. Pre-flight qualification, calibration and thermal testing were examined in Ref. 6.

This paper provides an overview of the CAMS and focuses specifically on in-flight performance. In the next section, an overview of the CAMS concept is provided. Commissioning activities are discussed in Section 3, and in-flight data analysis and processing is presented in Section 4. In-flight performance and results are discussed in Section 5.

## 2   CAMS Overview

*2.1 Basic Concept Design and Physical Characteristics*

The CAMS consists of two identical units. Each unit measures the lateral displacement (X/Y), and in combination, the rotation of the HXI plate relative to the fixed optical bench (FOB). The positioning of the CAMS units on the satellite is shown in Fig. 1. Each CAMS metrology system consists of a Laser and Detector module (CAMS-LD) and a Target module (CAMS-T).

The CAMS-LD modules are installed on the top plate of the FOB next to the HXTs. The CAMS-T is a passive retroreflector (corner cube mirror) placed on the HXI plate at the end of the EOB 12-meters away from the CAMS-LD.

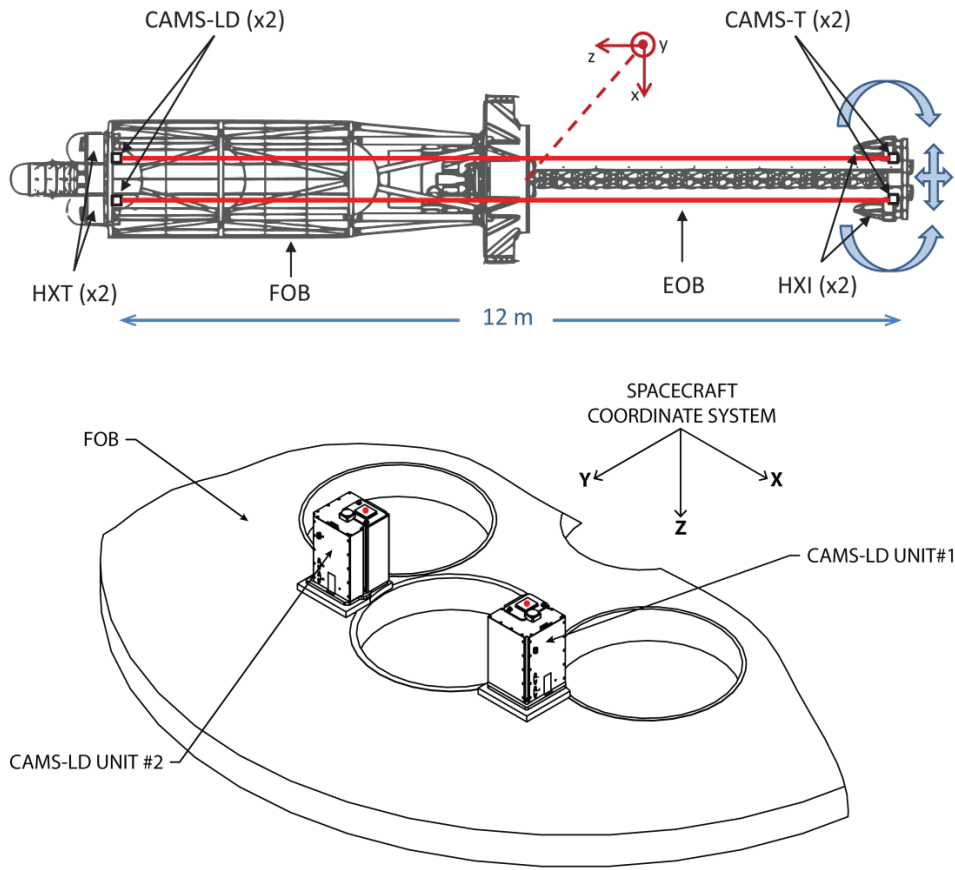

**Fig. 1** The location of the CAMS modules on the satellite is shown in the top figure. The positioning of the CAMS-LD on the top plate is shown in the bottom figure, the laser beam apertures are marked by red dots.

The CAMS concept is rather straightforward and depicted in Fig. 2. A continuous wave laser beam at 980 nm wavelength generated by a diode laser (3S Photonics 1994 SGP) is launched from the CAMS-LD modules and travels through the interior structure of the satellite. The laser strikes the retroreflector of the CAMS-T and is reflected back to the CAMS-LD. The beam expander in the LD unit reduces the beam size by an expansion ratio $M$. The CMOS imaging detector of 1024 x 1024 pixels detects the position of the laser beam. There is a linear relationship between the EOB lateral shift and the laser spot shift that is recorded by the CMOS detector. If the corner cube is displaced in the lateral direction, then the reflected laser spot shift is scaled by a factor of two. The measured shift from the nominal position on the CMOS detector will be scaled down accounting for the beam expander effects.

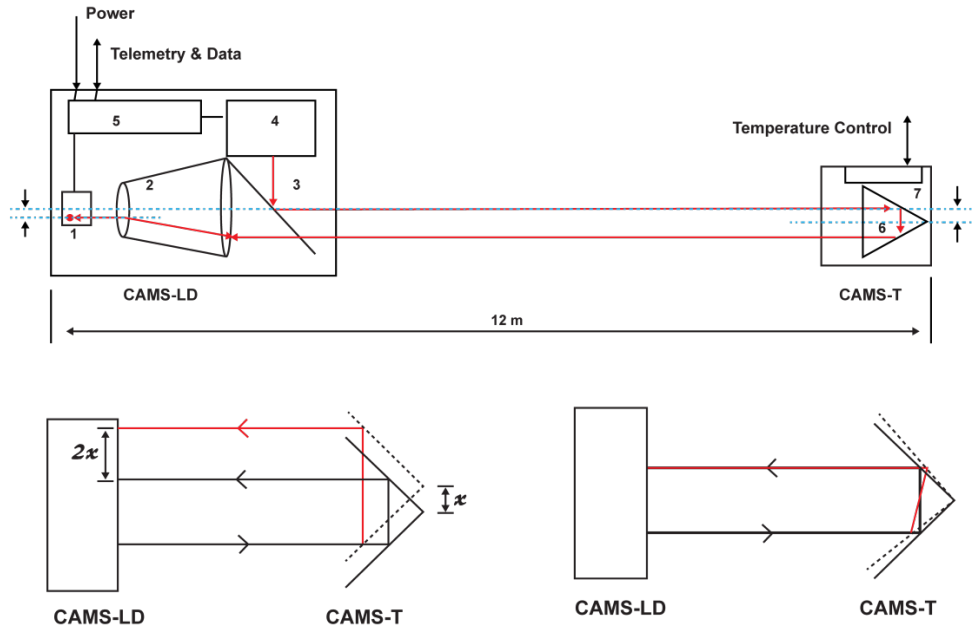

**Fig. 2** The top panel illustrates the path of the laser through the CAMS module. The laser is emitted from the collimator (4) then is deflected off a beamspliter (3) and directed toward the corner cube retroreflector (6). The beam is returned through the satellite interior, through the beamsplitter and into the scaling optics (2) before it is recorded on the CMOS detector (1). Other components of the system are the CAMS-LD electronics (5) and the CAMS-T heater and temperature sensor (7). The linearity of the system is depicted in the lower left. The robustness to tilt in the corner cube is illustrated in the lower right.

There are several inherent benefits to the CAMS design. The minimal divergence collimated laser beam makes the system less susceptible to background and stray light issues and minimizes concerns with stray laser light affecting other sensors on the satellite. The laser output power of several mW from the CAMS-LD, and the use of narrowband and neutral density filters, are important to overcome any expected solar background interference. In addition to the linearity of the system, it is also highly sensitive to lateral shifts. A 1 µm shift on the detector corresponds to a 2.4 µm shift of the EOB. The measurement is also not sensitive to the tilt of the corner cube around its apex (Fig. 2).  Physical characteristics of the CAMS system are provided in Table 1.

A description of the optical components and module structures can be found in Ref. 6. Images of the CAMS-LD and retroreflector are shown in Fig. 3.

Table 1 Physical characteristics of each CAMS unit.

| Attribute | Dimension |
| --- | --- |
| Field of view | 26 mm (diameter) |
| Sampling rate | 5 Hz |
| Mass | 3.3 kg (CAMS-LD) |
|  | 0.55 kg (CAMS-T) |
| Size | 165 x 165 x 200 mm (CAMS-LD) |
|  | 75 x 65 x 150 mm (CAMS-T) |
| Operating Temperature | -10 to +50 Celsius (CAMS-LD) |
|  | -20 to +75 Celsius (CAMS-T, function) |
|  | -10 to +65 Celsius (CAMS-T, performance) |
| Power | < 5 W (CAMS-LD) |
|  | < 12 W (CAMS-T, heater only) |
| Lifetime | > 3 years |

The key performance requirement for the CAMS system is the measurement accuracy of the HXI-plate position, which is set to a maximum $3\sigma$-error of 240 μm (or 4.297 arcsec).  The value is driven by the HXI detector pixel size and the expected point spread function of the HXT.  The accuracy requirement represents the key challenge for the CAMS development by imposing micro-radian scale pointing stability on the optical system while operating in a relatively wide temperature range. While there are other factors that contribute to the CAMS measurement error, their combined effect fall outside of the maximum allowance of 240 μm.  The most prevalent factor is the flexing of the FOB top plate onto which the CAMS-LD are mounted.

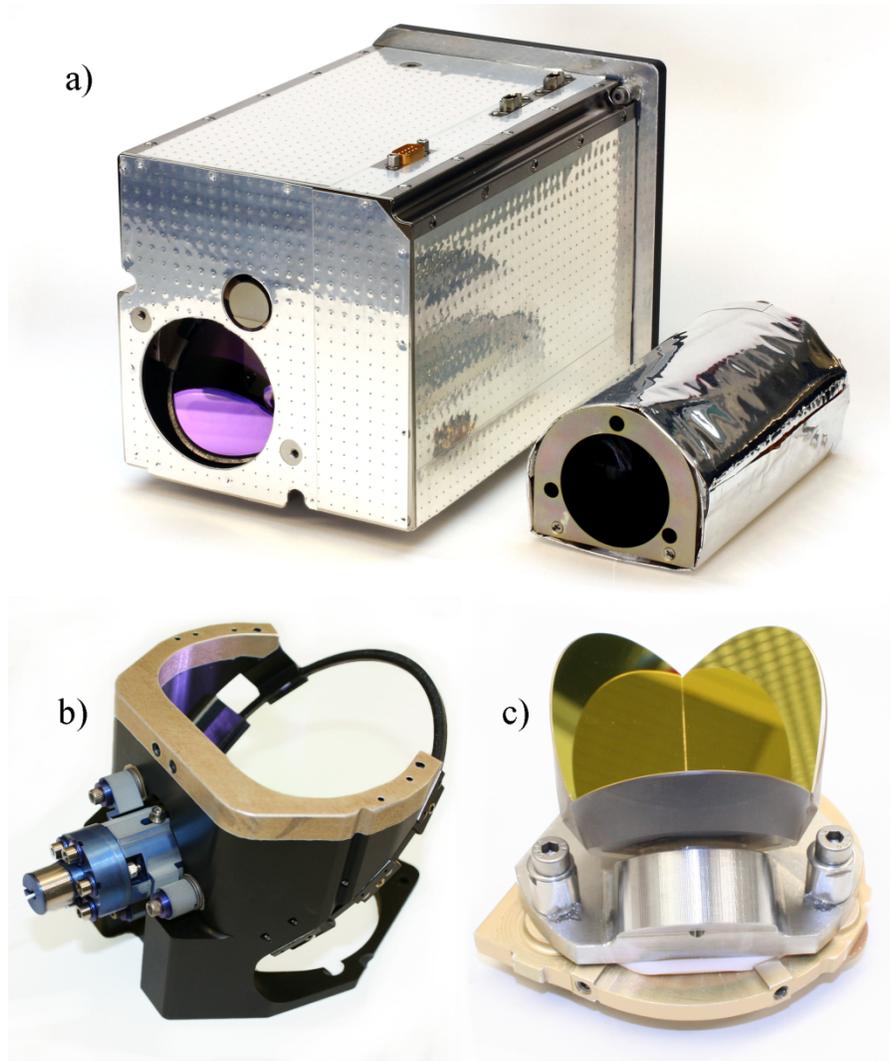

**Fig. 3** (a) The components, as delivered for integration with the spacecraft, in their respective housings (CAMS-LD on the left, CAMS-T on the right). Dimensions are given in Table 1. (b) A view of the critical optomechanical components of the CAMS-LD, beamsplitter and collimator assembly, (c) and of the CAMS-T, corner cube mirror.

*2.2 TVAC Performance Results*

CAMS data were collected throughout spacecraft-level Thermal Vacuum (TVAC) testing. Restricted by the size of the TVAC chamber, the EOB remained retracted as it would during launch. This provided a unique opportunity to measure the CAMS performance in the absence of flexing since the two modules were hard mounted between the two optical benches supporting them.

Solar lamps and shades were used to control the operational thermal set points. During testing, the shades were returned to the "closed" position in between set points. This created thermal creeping on the solar lamp exposed multi-layer insulation (MLI). This phenomenon creates stresses within the structure, which are observable in the CAMS data. Fig. 4 shows the measured structural distortion of about 12.6 μm for CAMS-1 and 4.6 μm for CAMS-2. The CAMS resolution is 1 μm with an rms measurement noise of approximately 0.5 μm.

The black dots superimposed on Fig. 4 relate to the detector raw images that were downloaded to verify that the measurements were not influenced by solar lamp background noise. The data are in the CAMS coordinate system and the units are rotated 90 degrees to each other on the FOB.

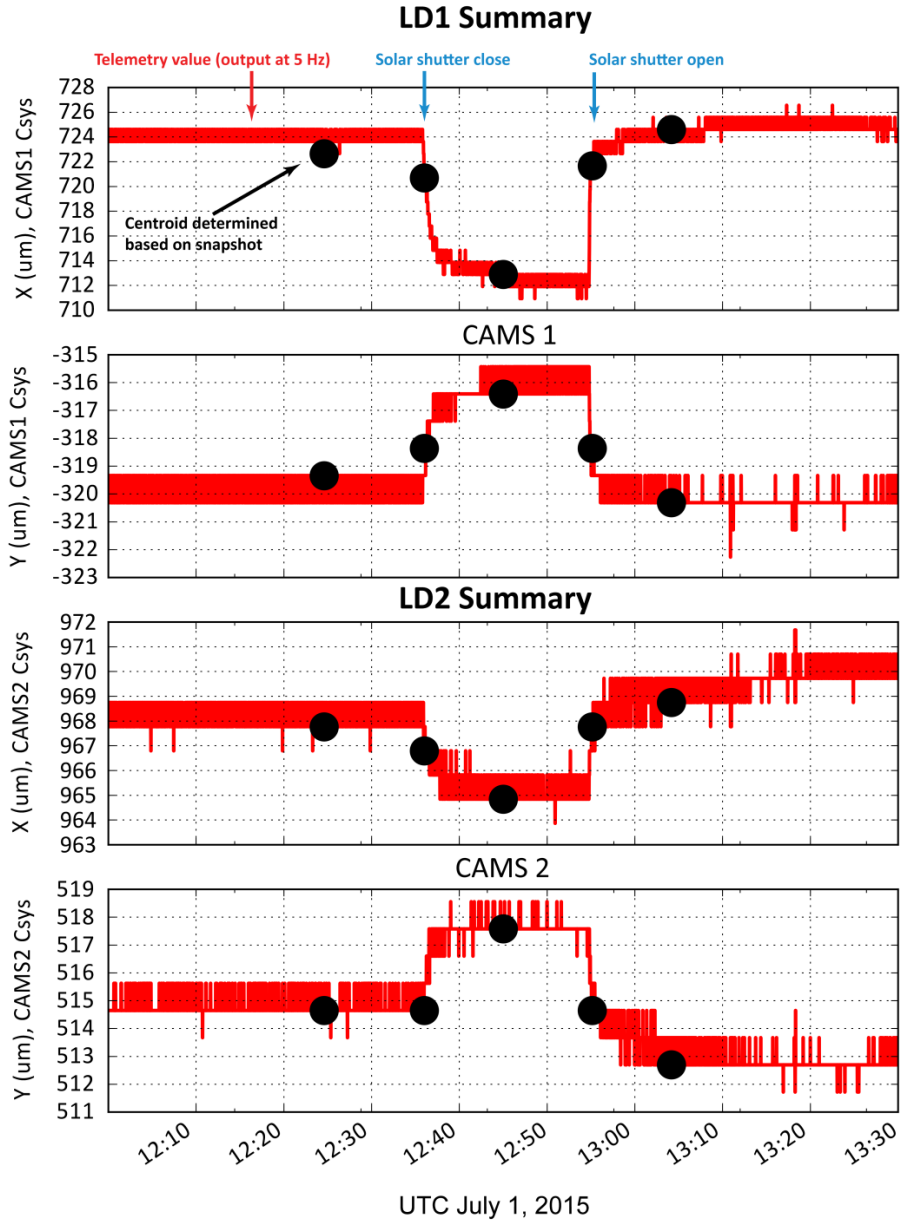

**Fig. 4** The spacecraft structural distortions measured in the X and Y direction by CAMS-1 (top two panels) and CAMS-2 (bottom two panels) during thermal vacuum testing of the spacecraft. The vertical axis shows the distortion in units of μm in the CAMS coordinate system (Csys). The black circles indicate the results of manual centroid calculations based on images downloaded from CAMS (full image dump).

*2.3 CAMS-LD Alignment on the FOB*

The alignment budget for differences between the incident angle of the laser emitted from the CAMS-LD and the satellite Z-axis were targeted to be less than 10 arcseconds in each lateral direction. The feasibility of achieving the alignment target by shimming the CAMS-LD was initially tested by JAXA with the CAMS engineering model during the Mechanical Interface Check in January 2014.

With the CAMS-LD flight models mounted on the top plate of the FOB, the laser spot position on the detector was measured when the CAMS-T was placed on the middle plate and then on the lower plate of the FOB. The top plate was 1.3 meters and 3.4 meters from the middle and lower plate, respectively. If the CAMS-LD alignment was perfect, then the laser spot would land on the same detector position when the CAMS-T was in either location. Shims were inserted between the CAMS-LD baseplate and the FOB top plate to achieve laser alignment.

After shimming, the laser orientation was measured to be within 15 arcseconds and 7 arcseconds in the X and Y lateral directions with respect to the satellite Z-axis for CAMS-1 and CAMS-2 units, respectively.

## 3  Commissioning Activities

The CAMS commissioning activities were performed over a 4-day period (Table 2) between February 27, 2016 (Launch + 10d) and March 1, 2016 (Launch + 13d). Commissioning activities occurred at the Uchinoura Space Center (USC) in the Satellite Telemeter Center, which houses a 34-m antenna, a control room and an equipment room for various devices such as transceivers. Commissioning uplinks and downlinks were also supported by Santiago Space Center (SNT).

**Table 2** CAMS/EOB operation summary during EOB deployment.

| Pass Date | Pass Ground Station | Commands |
|---|---|---|
| February 27 (Launch + 10d) | USC_1 | EOB-E turn on, status check, and motor heater control |
| | USC_2 | CAMS Power On, Laser On |
| | USC_3 | Temperature Read, CAMS-1 & -2 Snapshot; Download of image 1 |
| | USC_4 | Download of image 2 |
| February 28 (Launch + 11d) | USC_2 | HXI-HCE[1] off, HXI1-DPU[2] off, Non-Explosive Actuators release, EOB extension started (#1), stages 1 to 7 |
| | SNT1_1 | EOB extension (#2), stages 7 to 13 and then 13 to 19, AOCS[3] IRU[4] FDIR[5] threshold (increased from 0.07 deg/s to exceeding 0.2 deg/s) |
| | USC_3 | EOB extension (#3) stages 19 to 22, HXI1-DPU on, HXI-HCE on |
| | SNT1_2 | HXI1-DPU off, HXI-HCE off, EOB extension (#4) stages 22 to the latch point (full extension), HXI1-DPU on, HXI-HCE on |
| | USC_4 | AOCS RW1[6] on, Moment of Inertia parameter update (B), HXI2-DE[7], DPU on |
| | SNT1_3 | Moment of Inertia parameter update (A) |
| | USC5 | Temperature Read, CAMS-1 & -2 Snapshot; Download of image 1, NEAC[8] off, EOB-E off |
| | SNT1_4 | Download of image 2 |
| February 29 (Launch + 12d) | USC_3 | CAMS1.LASER_POWER 79; CAMS2.INTEGRATION_TIME 3; Temperature Read, CAMS-1 & -2 Snapshot; Download of image 1 |
| | SNT1_2 | Download of image 2 |
| | USC_4 | CAMS1.LASER_POWER 76; CAMS2.LASER_POWER 83; Temperature Read, CAMS-1 & -2 Snapshot; Download of image 1 |
| | SNT1_3 | Download of image 2 |
| | USC_5 | CAMS1.LASER_POWER 73; Temperature Read, CAMS-1 & -2 Snapshot; Download of image 1 |
| March 01 (Launch + 13d) | USC_2 | Download of image 2 |
| | USC_3 | Temperature Read, CAMS-1 & -2 Snapshot; Download of image 1 |
| | SNT1_3 | Download of image 2 - loss |
| | USC_4 | Temperature Read, CAMS-1 & -2 Snapshot; Download of image 1 |
| | USC_5 | Temperature Read, CAMS-1 & -2 Snapshot; Download of image 2, Download of image 1 - loss |

[1]- HXI-HCE – HXI Heater Control Electronics;
[2]- HXI-DPU – HXI Data Processing Unit;
[3]- AOCS – Attitude and Orbital Control System;
[4]- IRU – Inertial Reference Unit;
[5]- FDIR – Fault Detection Isolation and Reconfiguration;
[6]- RW1 – Reaction Wheel #1;
[7]- HXI-DE – HXI Central Processing Unit;
[8]- NEAC – Non-Explosive Actuator Controller.

*3.1 Turn-on and snapshots*

Activities consisted of reviewing the CAMS telemetry commands that needed to be sent to control the CAMS; reviewing telemetry to confirm the health of CAMS; and validating the image snapshot and download process in preparation for the next few days.

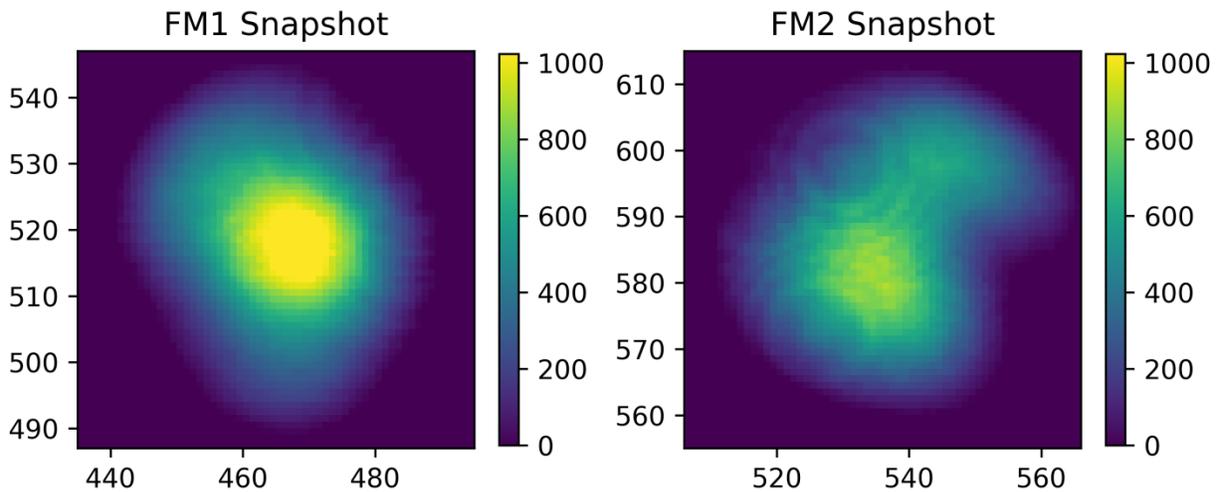

**Fig. 5** The imaged laser beam spot from CAMS-1 (left) and CAMS-2 (right) with the EOB stowed. The aberration in the images was expected as the EOB was not extended. In addition, the CAMS-1 image is saturated. Pixel values are shown on the axes. The color scale indicates relative intensity.

Figure 5 are the snapshot images taken on the day Launch +10 while the EOB was still stowed, showing that both units survived the launch. The distortions seen in the images were caused by laser beam segmentation on the corner cubes and diffraction on the corner cube center and boundaries between the petals. These distortions were expected to dissipate in the extended EOB configuration. Telemetry indicated the CAMS units were operating within nominal temperature range and there were no errors reported other than saturation of the unit readings in flight module-1 (FM1). This was fully anticipated and evident in the image (Fig. 5, left).

While the EOB was still stowed, the CAMS measurement deviations were compared with temperature changes in the telescope optics heaters and orbital temperature changes to investigate the influence of temperature variations on-board. Temperature changes in the heaters would occur on timescales of 10-20 minutes whereas orbital temperature fluctuations would occur on ~100 minute timescales. Fig. 6 illustrates how the heater cycling did indeed influence the CAMS measurements by inducing stresses in the FOB top plate onto which CAMS-LD was mounted. Note, the measurements in Fig. 6 were calculated assuming the extended length of the EOB to demonstrate the anticipated full impact of heater cycling on the CAMS measurements. Fluctuations in the displacement on the order of 1 mm are seen in the X and Y directions on timescales consistent with heater cycling. In the lower right panel of Fig. 6 the absolute accuracy of the system is tested. Here, the value *r12* is shown, which is the distance between the two units. The measurement is a sanity check to verify that both lasers travel in a parallel path, and nominally, the value of *r12* should be 600 mm. The effect of heater cycling introduced an uncorrectable error of approximately 80 μm rms, corresponding to 1.5 arcsec, and will be further discussed in Section 3.4.

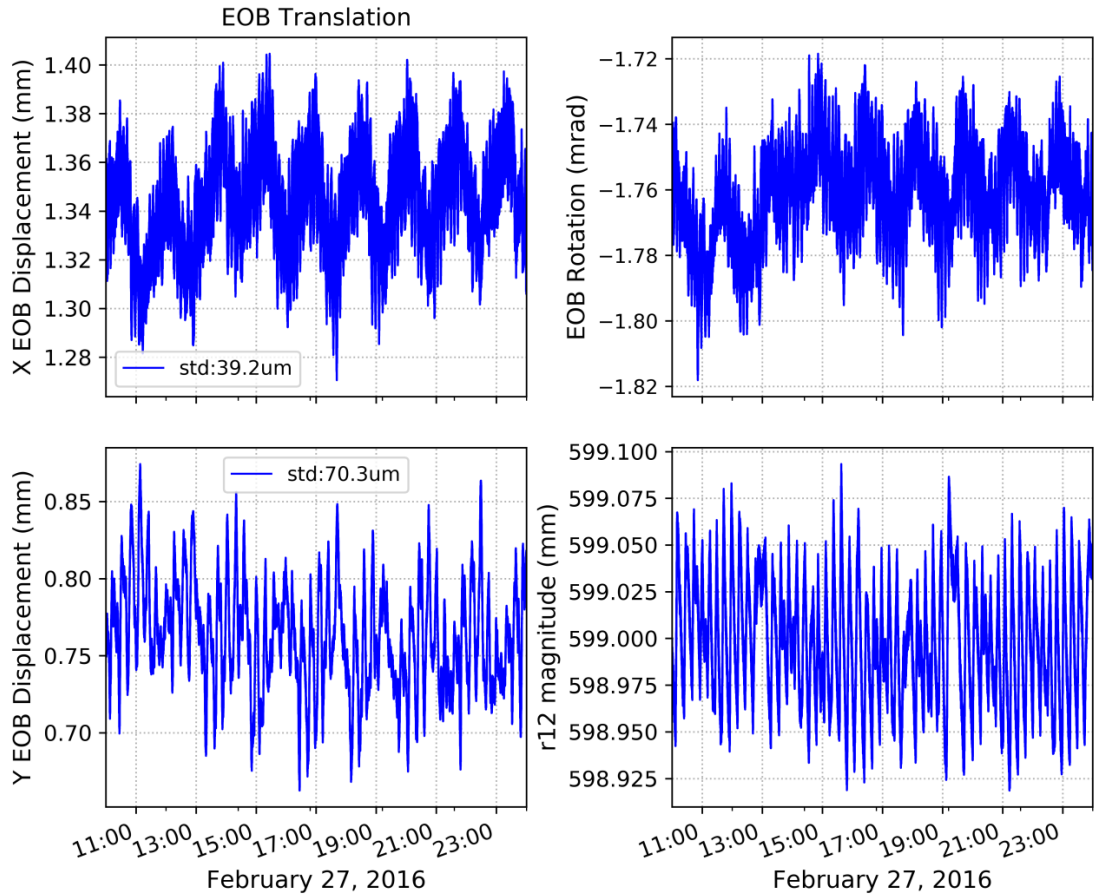

**Fig. 6** The effect of the telescope optics heater cycle on the CAMS measurements of the EOB position. Lateral translation in *X* (top left) and *Y* (lower left), rotation around *Z* axis (top right) and the distance measured between CAMS-T1 and CAMS-T2 (*r12*, design value is 600 mm) (lower right) indicating the absolute accuracy. The horizontal axes show timescales in hh:mm format (UTC). A one-mm shift corresponds to 17.19 arcsec angular shift in the focal plane (at HXI) when the EOB is extended.

*3.2 EOB Extension*

On day Launch + 11, the CAMS was used to monitor the deployment of the EOB[7]. The EOB is an extensible mast structure comprising of 23 stages, of which 22 are extensible. After full extension, it becomes 6377 mm in length (689 mm in stowed configuration). The EOB, whose total weight is 42.2 kg, pushes and sustains the HXI plate, which has a mass of approximately 150 kg, throughout the extension operation. During EOB extension, the CAMS generated X/Y data pairs with both units. At the same time, the triaxial angular velocities of the spacecraft were monitored with the Inertial Reference Unit (IRU).

Figure 7 shows the calculated EOB translation and rotation based on the combined data from both units. The extension motion can easily be unstable even for small perturbations since the HXI plate is massive and the EOB is not completely stiff. In fact, the HXI plate was sufficiently unstable in the lateral direction that the extension operation was manually suspended on four occasions that were determined by monitoring the CAMS and the IRU data. These delays occurred at 2:17UT, 2:59UT, 3:02UT and 3:53UT during the extension of the $7^{th}$, $14^{th}$, $20^{th}$ and $23^{rd}$ stages, respectively. The vibrations damped quickly (within ~30 seconds) and the extension process resumed after confirmation that the EOB was sufficiently steady. Even with such careful consideration, the unsteadiness was unexpectedly large during the last extension to the latch point that the rotation rate of one of the four reaction wheels (RWs) exceeded a pre-set limit and that RW was shut off by the Fault Detection Isolation and Reconfiguration (FDIR). Despite such challenges, the EOB was fully extended and the CAMS real-time data proved extremely beneficial in the process.

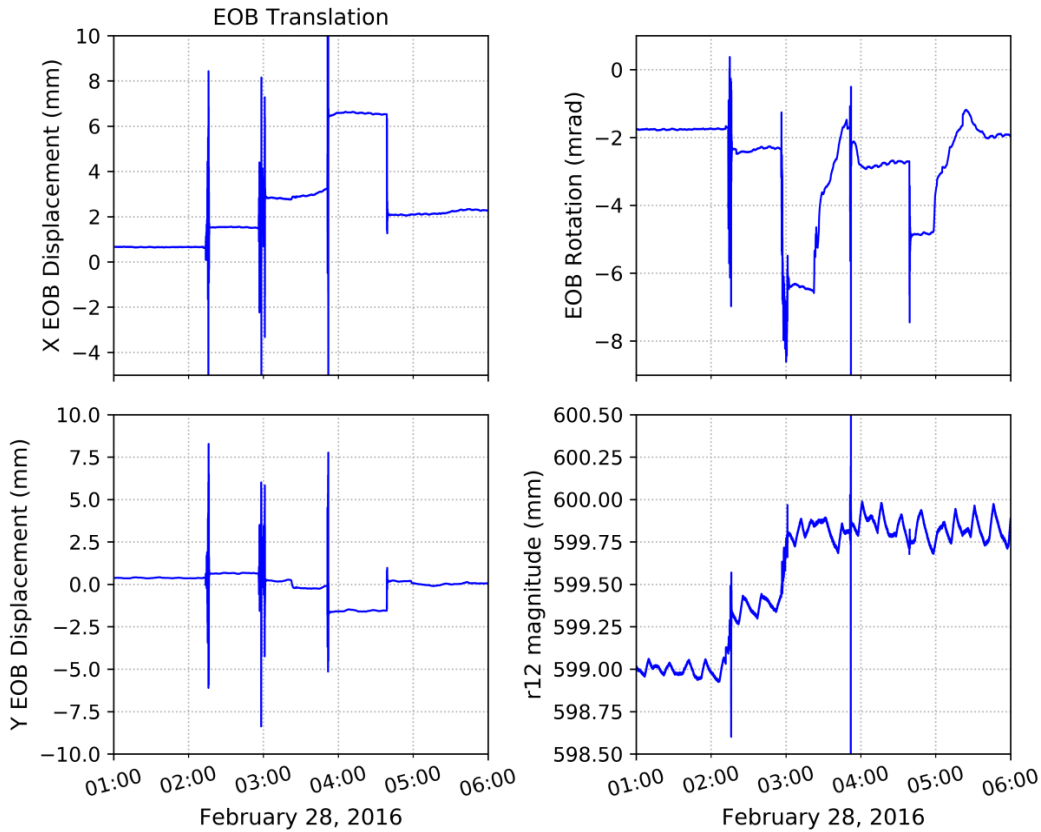

**Fig. 7** EOB calculated translation and rotation based on CAMS measurements during the entire period of EOB deployment (horizontal axis is time). Lateral translation in *X* (top left) and *Y* (lower left), rotation around *Z* axis (top right) and the distance measured between CAMS-T1 and CAMS-T2 (*r12*, design value is 600 mm) (lower right) indicating the absolute accuracy. The horizontal axes show timescales in hh:mm format (UTC). A one-mm shift corresponds to 17.19 arcsec angular shift in the focal plane (at HXI) when the EOB is extended.

*3.3 Parameter Adjustment*

On day Launch + 12, after successful EOB deployment, the snapshots in Fig. 8 were obtained and inspected to ensure the laser beams were operating normally. The laser beam shapes were Gaussian, as expected, alleviating concerns prior to launch that interference patterns were observed from both units. The concentric patterns observed in Fig. 8 are effects of laser diffraction on dust particles present in the optical system close to the CAMS detector. This diffraction does not impact the CAMS centroid calculation and measurement. A beam quality factor, returned in the telemetry, matched values obtained during the unit calibration process. The beam quality factor provided by CAMS was used to quantify the quality of the data that goes into calculating the centroid. It is based on the Pearson Coefficient to compare sampled data points to a Gaussian distribution representing the expected laser beam shape on the detector.

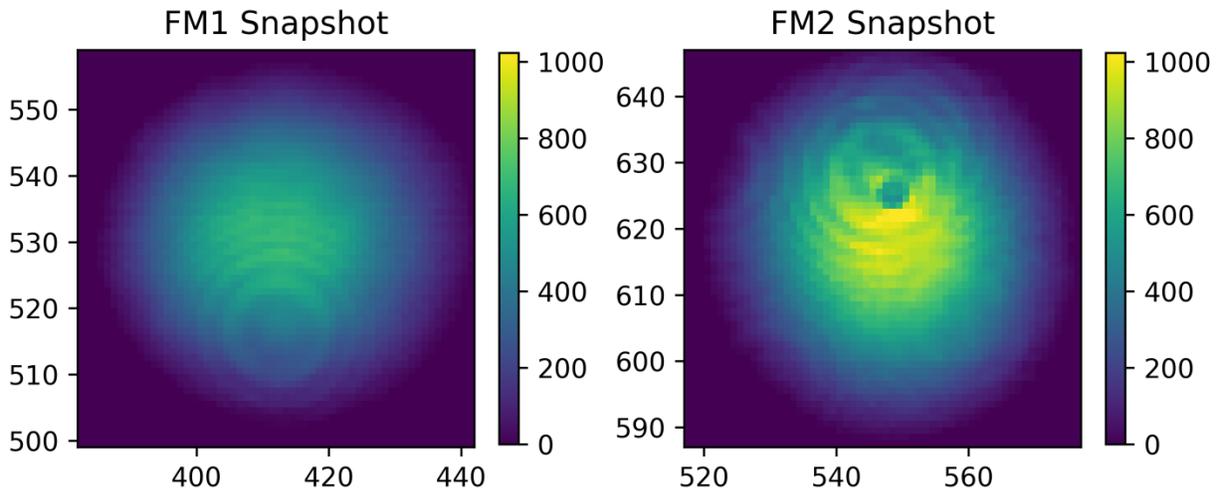

**Fig. 8** The CAMS beam shape from each unit taken after the EOB deployment. Pixel values are shown on the axes. The color scale indicates relative intensity.

Based on ground calibration data of the expected return intensity over the detector (Fig. 9, right panel) and parameter sensitivity, the detector gain and laser current could be sequentially modified until reaching the desired intensity (blue dots in Fig. 9, left panel). Initially, the returning laser intensity (red dots Fig. 9, left panel) was slightly higher compared to the calibration levels. The returning laser intensity was adjusted until it was within the calibration range (i.e. the red dot fell within the blue uncertainty range in Fig. 9).

The right panel in Fig. 9 shows the expected intensity returned as a function of beam position on the detector. Fig. 9 (right) indicates the final beam position (white cross) and the beam size (magenta region concentric to the white cross) relative to the detector FOV.

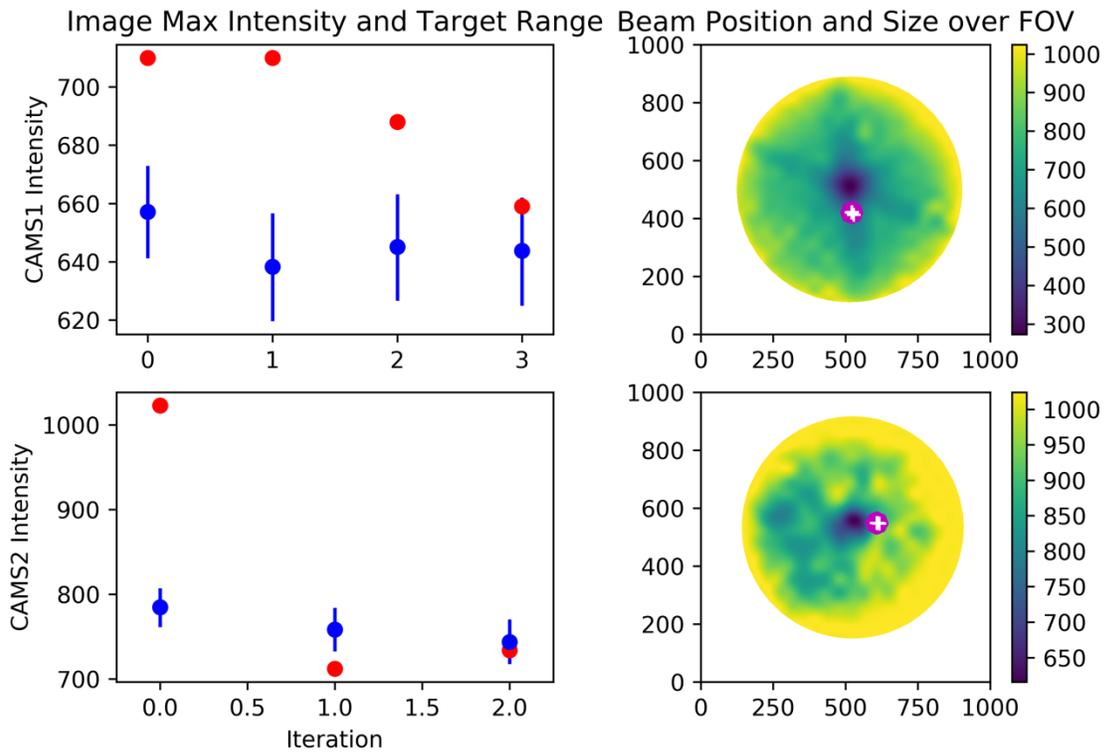

**Fig. 9** *Left*: Parameters Adjustment Sequence for CAMS-1 (top) and CAMS-2 (bottom). The red dots indicate the CAMS telemetry measurement of the total intensity parameter. The blue dots are the calibration levels for the current beam location on the FOV with the error bars. *Right*: The final beam position (white cross) and the beam size (magenta region concentric to the white cross) over the expected intensity distribution maps measured during ground calibration (on the right).

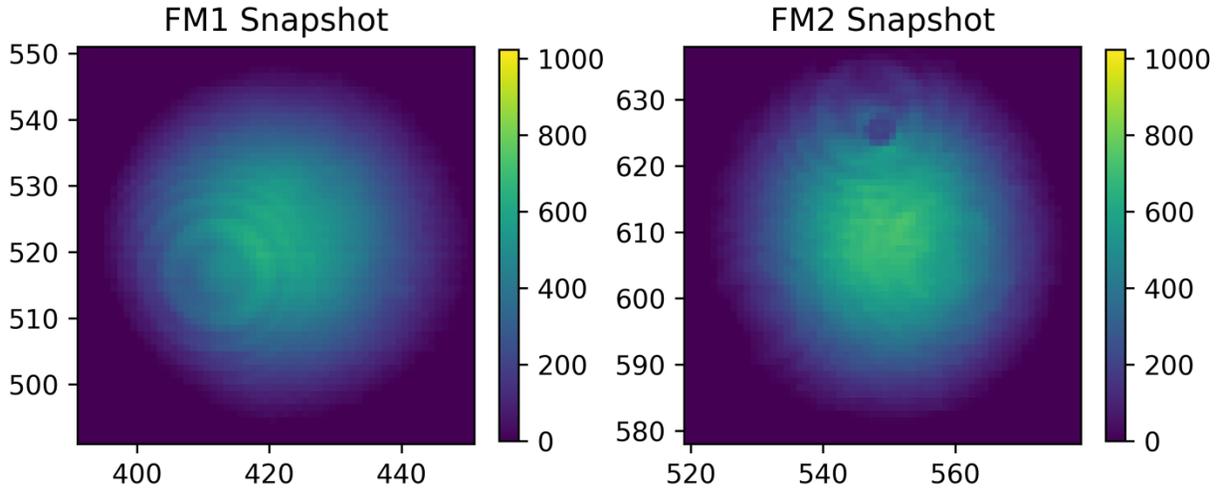

**Fig. 10** The final images acquired by CAMS showing Gaussian laser beams of similar intensity in the CAMS-LD1 (left) and CAMS- LD2 (right). Pixel values are shown on the axes. The color scale indicates relative intensity.

The final calibrated image is displayed in Fig. 10 and shows a Gaussian laser beam of similar intensity in each unit.

*3.4  Orbital effects*

The CAMS parameter adjustment activities were successfully concluded on Launch + 13 days. Snapshots continued to be acquired from both units to evaluate the parameters throughout the day. Due to minimal EOB motion, the same detector area was consistently imaged. Laser beam quality remained nominal throughout the day providing confidence in the parameter selection.

As CAMS telemetry data were cumulated, the effects of night/day cycles on the spacecraft structure became apparent. As shown on Fig. 11, the effect is mostly noticeable in the EOB rotation around the Z-axis (Fig. 11, upper right) where there is clear delineation between night and day. A sinusoidal oscillation of ~175 µm (3 arcsec) peak-to-peak amplitude can be seen in the X direction (upper left). The Y direction exhibits similar oscillations, but is more uneven by what is attributed to the FOB distortions induced by the optics heater cycles. The bottom right panel in Fig. 11 shows the calculated distance between the two measurements (*r12*), which should nominally be 600 mm if both lasers travel in an ideal parallel path. After deployment of the EOB, the average value was 600.62 mm indicating very small deviations from the projected, gravity-free, FOB deflections. The variations of this value are due to slight differences in the laser direction from the two CAMS units due to thermal distortion of the FOB top plate.

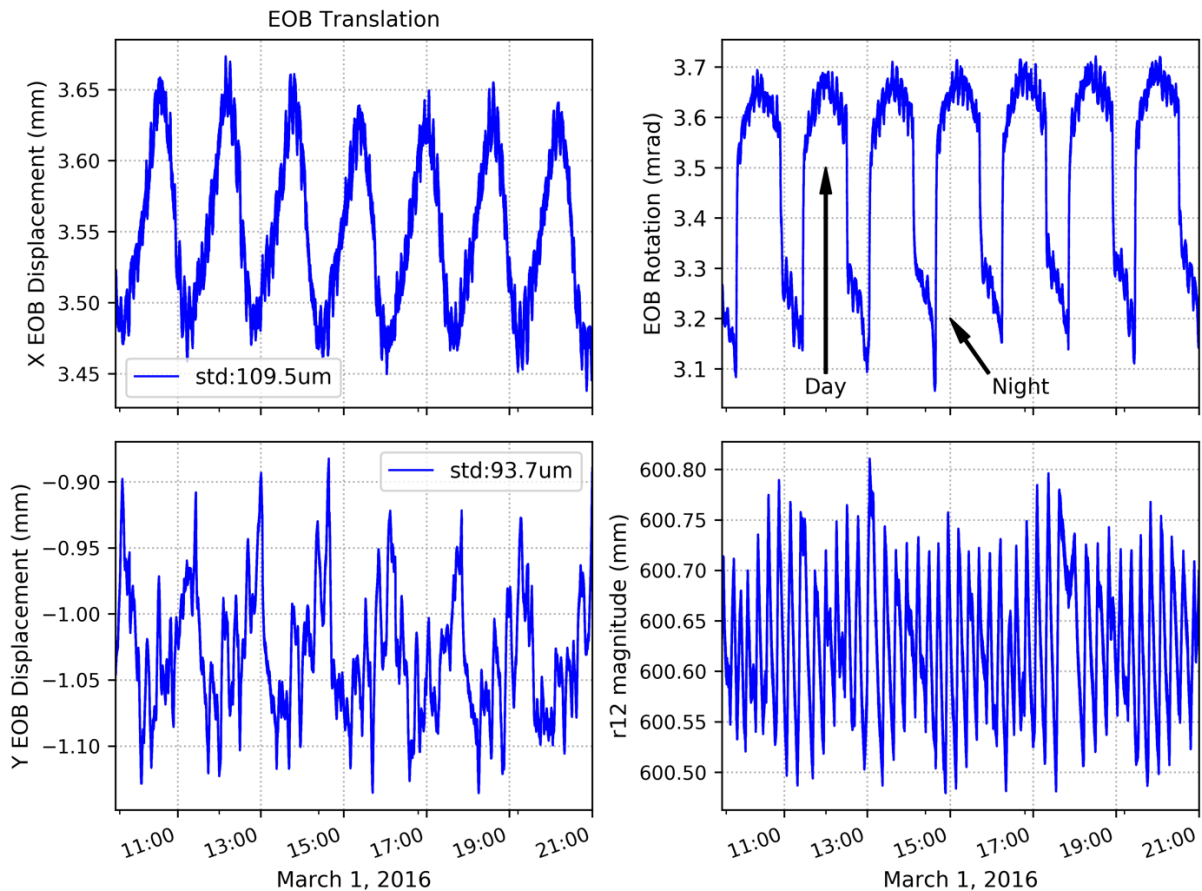

**Fig. 11** Calculated translation and rotation in the EOB based on CAMS measurements during day and night cycles. Lateral translation in *X* (top left) and *Y* (lower left), rotation around *Z* axis (top right) and the distance measured between CAMS-T1 and CAMS-T2 (*r12*, design value is 600 mm) (lower right) indicating the absolute accuracy. The horizontal axes show timescales in hh:mm format (UTC). A one-mm shift corresponds to 17.19 arcsec angular shift in the focal plane (at HXI) when the EOB is extended.

Data were scrutinized to explain the higher frequency excursions measured by CAMS. By comparing with the telescope optics heater commands (Fig. 12), correlations with the CAMS data could be confirmed. Heater cycling introduces heat on the FOB top plate, creating stresses, which alter the CAMS-LD laser launch angle. These changes in the launch angle are measured as an EOB deflection on the order of ~100 μm. The effects of heater cycles ultimately limit the accuracy of the CAMS measurements. Although the correlation with heater cycles is evident, it is difficult to accurately distinguish heater induced displacement in laser launch angle (error) from actual EOB motions.

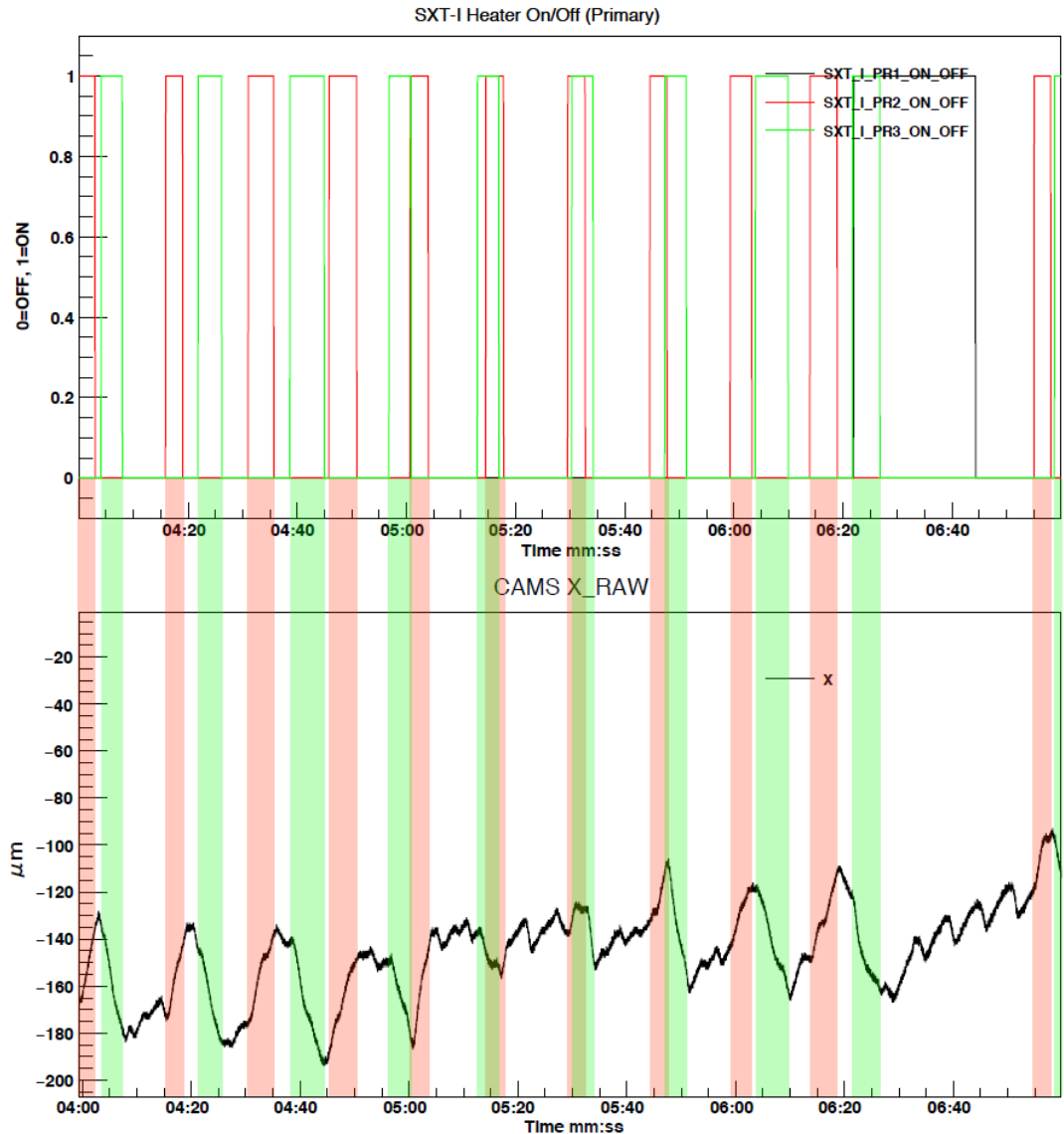

**Fig. 12** Comparison of CAMS-1 raw data in the X direction and the Soft X-ray Telescope heaters over the same time frame.

## 4   In-Flight Data Analysis

*4.1 HXI astronomical flight data*

During the limited lifetime of the mission there were two HXI observations of astronomical objects conducted where the CAMS corrections could be applied and its importance validated. The X-ray observations were of the Crab and the supernova remnant G21.5-0.9. These observations were of significant interest from an instrument calibration point-of-view as the energy distribution in the image is spherical and peaked at the center. The distribution remains constant even when observations are carried out over long periods, therefore any variation would be attributed to spacecraft pointing inaccuracy or movement within the telescope structure.

X-ray observations are composed of discrete photon events associated with a pixel position on the imager and a timestamp for an effective exposure of several hours. JAXA had pre-processed the two datasets to contain average pixel position, pixel count, and spacecraft attitude within a sampling period. When aggregated together and corrected for motion, the images can be reconstructed as shown in Fig. 13.

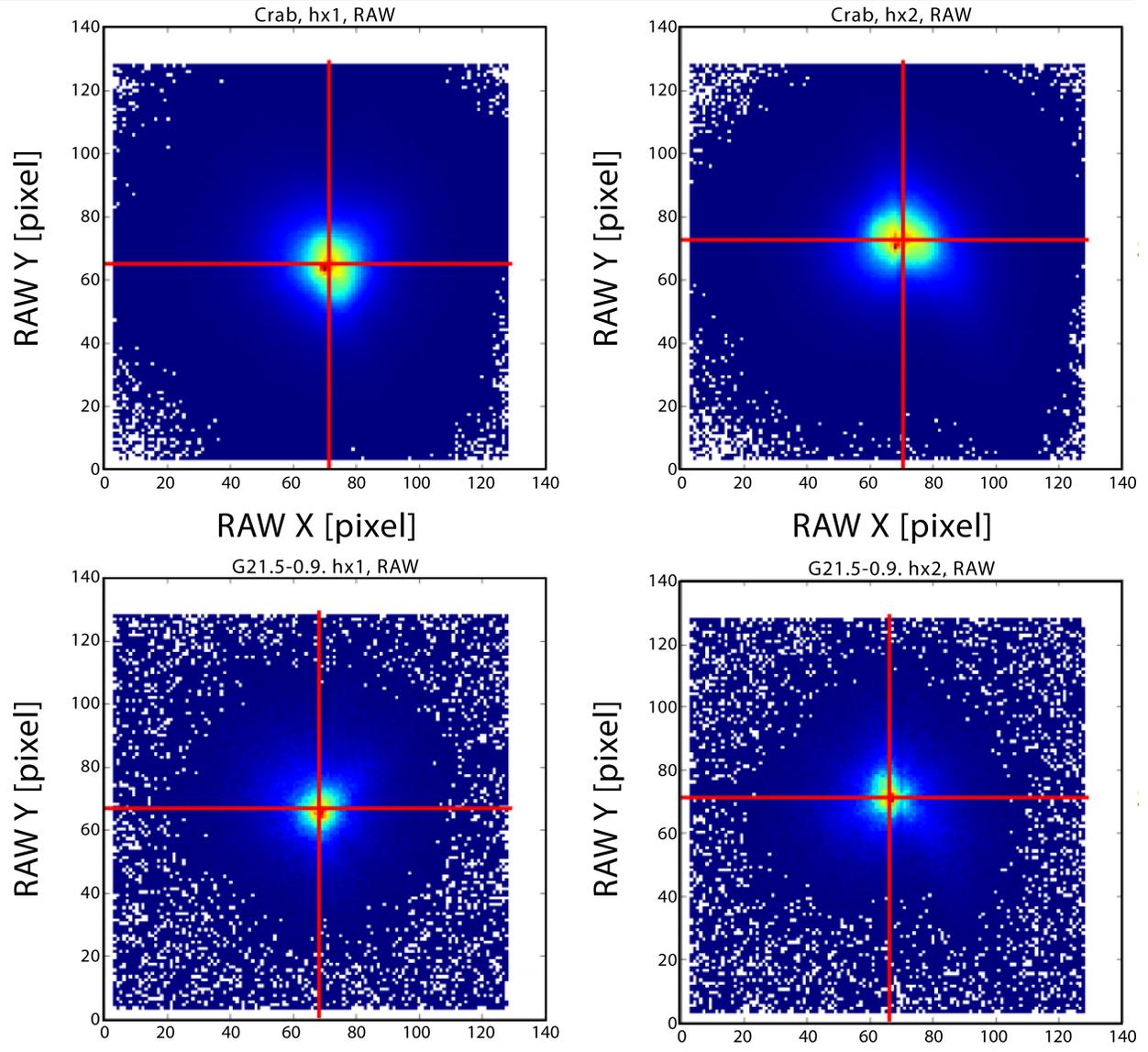

**Fig. 13** Raw HXI-1 (left) and HXI-2 (right) images of the Crab (top) and G21.5-0.9 (bottom). The FOV of HXI sensors corresponds to 9.17 arcmin in angular scale.

Table 3 shows the sampling period for each X-ray source and the average number of photons received. The count rate is significantly lower for G21.5-0.9 resulting in a longer sampling period (hence more time for telescope fluctuations), more scatter, and a greater number of outliers in the centroid measurements. Photons were gathered in the HXI 1/2 coordinate frames, which are rotated +/-22.5° to the spacecraft coordinate system (Fig. 15).

Table 3 X-ray point source data.

|  | Sampling Period | Mean Photon Count / Period | Photon Count Std / Period |
|---|---|---|---|
| **Crab** | 1 min | 19,671 | 3,048 |
| **G21.5-0.9** | 5 min | 417 | 145 |

*4.2 Flight data processing*

This section describes the methodology employed in using CAMS data to correct the HXI images. The procedure is based on earlier work[5] and includes several updates and improvements.

The orientation of the spacecraft body-frame, denoted by SAT, is shown in Fig. 14 along with the main spacecraft structure. Fig. 14 identifies the location and orientation of the coordinate frames for both units of the CAMS and HXI. The CAMS measurements are reported in the CAMS-1 and CAMS-2 reference frames. The orientations of these local coordinate frames are defined by the orientation of each alignment cube,[6] but their origins are the geometrical center of the FOV of each unit.

A set of CAMS alignment parameters is required for determining the transformation from the CAMS coordinates to the satellite coordinate system (Table 4).

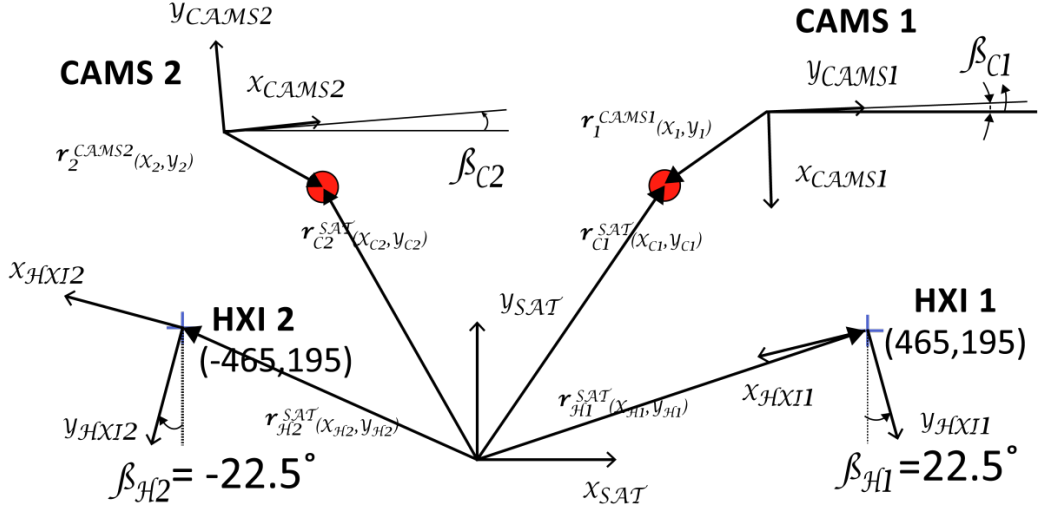

**Fig. 14** CAMS and HXI coordinate frames as viewed looking from the FOB along the EOB. The nominal (as designed) location of each instrument is given in mm as a coordinate pair $(x, y)$ in the SAT coordinate frame, except for the CAMS readings which are in the CAMS local coordinate frames. Red circles represent the center of CAMS-T1 and CAMS-T2 in an undisturbed EOB location. The origin in the CAMS frames are the centers of their field-of-view.

Table 4. CAMS installation parameters.

| Description | Notation | Values[1] |
|---|---|---|
| Spacecraft coordinates of the center of the CAMS-T1 unit | $\mathbf{r}_{C1}^{SAT}(x_{C1}, y_{C1}, z_{C1})$ | (300,480,-5723) |
| CAMS-LD1 Alignment cube rotation angle about its z axis, deg | $\beta_{C1}$ | -0.674 |
| CAMS1 Reported values at zero EOB displacement | $\mathbf{r}_1^{CAMS1}(x_1, y_1)$ | (-0.254,1.133) |
| Spacecraft coordinates of the center of the CAMS-T2 unit | $\mathbf{r}_{C2}^{SAT}(x_{C2}, y_{C2}, z_{C2})$ | (-300,480,-5723) |
| CAMS-LD2 Alignment cube rotation angle about its z axis, deg | $\beta_{C2}$ | +0.155 |
| CAMS2 Reported values at zero EOB displacement | $\mathbf{r}_2^{CAMS2}(x_2, y_2)$ | (0.879,2.246) |
| HXI-1 Reference frame origin in SAT coordinate frame | $\mathbf{r}_{H1}^{SAT}(x_{H1}, y_{H1}, z_{H1})$ | (465,195,-5491) |
| HXI-1 Reference frame rotation, deg | $\beta_{H1}$ | 22.5 |
| HXI-2 Reference frame origin in SAT coordinate frame | $\mathbf{r}_{H2}^{SAT}(x_{H2}, y_{H2}, z_{H2})$ | (-465,195,-5491) |
| HXI-2 Reference frame rotation, deg | $\beta_{H2}$ | -22.5 |

[1] – Coordinate values are displayed in mm.

Two other frames related to the HXI that are relevant to the data processing are the RAW and ACT coordinate frames (Fig. 15). The ACT (short for active) coordinate frame is a virtual coordinate frame representing the ideal (or nominal) location of the HXI, that is, the location of the HXI when all deformations of the EOB are zero. The RAW coordinate frame is the base sensor frame attached to the HXI. The raw position of photon events observed by the imager are expressed in RAW coordinates, and as such, the RAW frame translates and rotates along with the HXI detectors mounted on the end of the EOB. In the case of zero EOB deformation the ACT, RAW, and HXI frames all have the same orientation but their origins are offset.

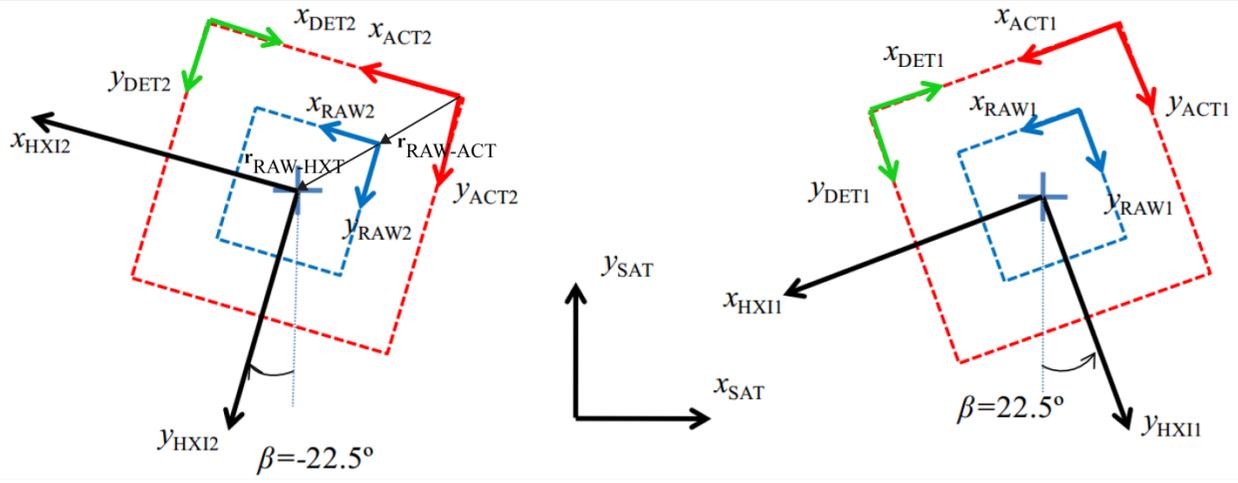

**Fig. 15** HXI reference frame to spacecraft reference frame.

Prior to applying a CAMS correction to HXI measurements, it is important to ensure that an adequate correction of thermal pointing drifts is applied to the CAMS data themselves. This correction takes into consideration the laser beam displacement caused by temperature variation in the CAMS. Ground based unit level tests determined linear thermal drift coefficients for correcting the CAMS internal contribution. A contribution to thermally induced laser pointing drift caused by the top plate bending is not included in this correction. The options to implement such a correction were either (i) though on-board processing or (ii) during ground processing. The ground based approach was selected thus the correction became part of the data processing.

The temperature corrected CAMS readings can be calculated using the following pair of expressions:

$$x_{Corrected1,2} = x_{CAMS1,2} - \Delta_{x1,2}^{CAMS}(T_{CAMS1,2} - T_{Calibration1,2})$$
$$y_{Corrected1,2} = y_{CAMS1,2} - \Delta_{y1,2}^{CAMS}(T_{CAMS1,2} - T_{Calibration1,2}) \qquad (1)$$

where for each CAMS unit 1 and 2, $x_{CAMS}$ and $y_{CAMS}$ are CAMS readings in $x$ and $y$; $\Delta_x^{CAMS}$ and $\Delta_y^{CAMS}$ CAMS are thermal coefficients obtained via unit level thermal tests; $T_{CAMS}$ is the temperature of the optical assembly within CAMS and reported in the CAMS telemetry; and $T_{Calibration}$ is the temperature of the CAMS optics during the ground calibration of the unit. In this case, only one temperature sensor is required for the CAMS correction. Table 5 lists the calibration values required for the thermal correction.

Table 5 Thermal correction parameters for CAMS.

| Description | Notation | Values |
|---|---|---|
| CAMS1 thermal coefficient in x, mm/deg C | $\Delta_{x1}^{CAMS}$ | -0.021 |
| CAMS1 thermal coefficient in y, mm/deg C | $\Delta_{y1}^{CAMS}$ | 0.0262 |
| CAMS1 ground calibration temperature, deg C | $T_{Calibration1}$ | 22.8 |
| CAMS2 thermal coefficient in x, mm/deg C | $\Delta_{x2}^{CAMS}$ | -0.0199 |
| CAMS2 thermal coefficient in y, mm/deg C | $\Delta_{y2}^{CAMS}$ | 0.0364 |
| CAMS2 ground calibration temperature, deg C | $T_{Calibration2}$ | 23.1 |

Next, the four CAMS measurements are used to estimate three parameters representing the three significant relative motion degrees-of-freedom: the 2-D planar translation and the rotation about an axis parallel to the boresight. The data processing algorithm begins by calculating the twist angle (*γ*) followed by the planar displacement. Although the CAMS measurements are planar, they are represented as 3-D position vectors with a zero value in the z-direction, i.e. $\mathbf{r}_1^{CAMS1} = [x_1 \quad y_1 \quad 0]$ and $\mathbf{r}_2^{CAMS2} = [x_2 \quad y_2 \quad 0]$ in each local CAMS coordinate frame (denoted with a superscript). The measurement vectors and the twist angle are shown in Fig. 16, along with the original position of the reflected laser beams 1 and 2, and the displaced position denoted with a prime superscript. The position vectors from one laser beam location to the other, $\mathbf{r}_{12}$ and $\mathbf{r}'_{12}$ are combined with a dot product to calculate the twist angle:

$$\gamma = \cos^{-1}\left(\frac{\mathbf{r}_{12} \cdot \mathbf{r}'_{12}}{|\mathbf{r}_{12}||\mathbf{r}'_{12}|}\right) \quad (2)$$

To determine the sign of the twist angle the cross product of $\mathbf{r}_{12}$ and $\mathbf{r}'_{12}$ is used:

$$\text{sign}(\gamma) = \text{sign}(\mathbf{r}_{12} \times \mathbf{r}'_{12}) \quad (3)$$

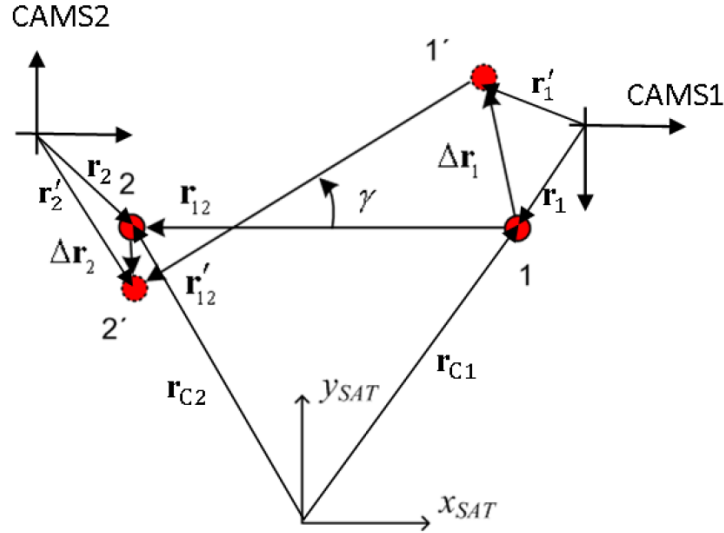

**Fig. 16** Estimation of roll angle, $\gamma$, using the planar displacement of the two CAMS-T.

The previous equations assume that the position vectors $\mathbf{r}_{12}$ and $\mathbf{r}'_{12}$ are known and expressed relative to a common reference frame. To calculate $\mathbf{r}_{12}$ and $\mathbf{r}'_{12}$ the location of each CAMS-T unit, $\mathbf{r}_{C1}$ and $\mathbf{r}_{C2}$ must be known (see Table 4) and the CAMS measurements, $\mathbf{r}_1$ and $\mathbf{r}_2$, must be expressed in a common reference:

$$\mathbf{r}_{12} = \mathbf{r}_{C2} - \mathbf{r}_{C1} \tag{4}$$

$$\mathbf{r}'_{12} = \mathbf{r}_{C2} + \mathbf{r}'^{SAT2}_2 - \mathbf{r}^{SAT2}_2 - (\mathbf{r}_{C1} + \mathbf{r}'^{SAT1}_1 - \mathbf{r}^{SAT1}_1)$$

$$\Delta \mathbf{r}_1 = \mathbf{r}'_1 - \mathbf{r}_1$$

$$\Delta \mathbf{r}_2 = \mathbf{r}'_2 - \mathbf{r}_2$$

It is assumed here that the original reference orientation of the EOB is the non-disturbed nominal orientation, where the targets are located at nominal coordinates. The shifted reading of CAMS must consider a bias ($\mathbf{r}^{CAMS1,2}_{1,2}$) measured at spacecraft calibration (Table 4). A rotation is necessary to express the CAMS measurements in a common frame such as the SAT frame:

$$\mathbf{r}^{SAT}_1 = R^{SAT}_{CAMS1} \mathbf{r}^{CAMS1}_1$$

$$\mathbf{r}^{SAT}_2 = R^{SAT}_{CAMS2} \mathbf{r}^{CAMS2}_2 \tag{5}$$

For the orientation of the frames in Fig. 16, the following rotations are used:

$$R^{SAT}_{CAMS1} = R_z(-\frac{\pi}{2} + \beta_{C1})$$

$$R^{SAT}_{CAMS2} = R_z(\beta_{C2}) \tag{6}$$

where $R_z(\theta)$ is a 3-D rotation matrix for a rotation of an angle $\theta$ about the z-axis.

Once the twist angle is determined, the planar displacement of the EOB is estimated using the centroid position of the CAMS-T1,2 locations. The displacement ($\Delta \mathbf{r}$) is the difference between a vector pointing to the centroid of the original CAMS-T1,2 locations, $\mathbf{r}_{c/m}$, after it has been rotated by the twist angle, $\gamma$, and a vector to the centroid of the displaced CAMS-T1,2 locations, $\mathbf{r}'_{c/m}$:

$$\Delta \mathbf{r} = \begin{bmatrix} \Delta x \\ \Delta y \\ 0 \end{bmatrix} = \frac{z_{H1}}{z_{C1}} \left( \mathbf{r}'_{c/m} - R_z(\gamma) \mathbf{r}_{c/m} \right) \tag{7}$$

This step is shown graphically in Fig. 17.

Here a vector pointing to the centroid of any two vectors, $\mathbf{r}_1$ and $\mathbf{r}_2$, is found using the following relationship:

$$\mathbf{r}_{c/m} = \mathbf{r}_1 + \frac{1}{2}(\mathbf{r}_2 - \mathbf{r}_1) \tag{8}$$

Expression (6) also takes into consideration that the HXI detectors are elevated 232 mm above the HXI plate and the lateral translational displacement measured by CAMS must be factored. It is found that the higher the HXI detectors are located, the smaller the motion will be if it is caused by bending of the EOB. This does not hold true for the lateral motion associated with the twist angle of the EOB. The rotation associated with twist is preserved without an extra factor. This consideration also means that calculating the HXI detector displacement $\Delta \mathbf{r}$ in the reference frame centered in the physical axis of the rotation becomes important.

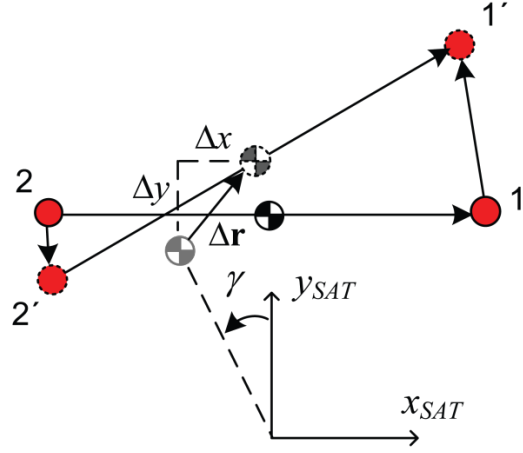

**Fig. 17** Illustration of the planar translation, Δr, using the centroid of the original and displaced CAMS-T1,2 locations.

Once the CAMS data are processed using equations (1) to (8), the position of photon events observed by the HXI, $\mathbf{r}_{RAW}$, can be corrected to account for the measured distortion of the EOB relative to the FOB. For this algorithm, the data $\mathbf{r}_{RAW}$ already includes any required correction from spacecraft attitude control sensors.

The corrected HXI observation in active coordinates, $\mathbf{r}_{ACT}$, are calculated directly using the calculated displacement vector, Δr, and twist angle, γ, using the following expression:

$$\mathbf{r}_{ACT} = \mathbf{r}_{ACT-RAW} + R_{RAW}^{ACT}\mathbf{r}_{RAW} + R_{SAT}^{ACT}[R_z(\gamma)\mathbf{r}_{0RAW} + \Delta\mathbf{r} - \mathbf{r}_{0RAW}] \qquad (9)$$

where $R_z(\gamma)\mathbf{r}_{0RAW} + \Delta\mathbf{r} - \mathbf{r}_{0RAW}$ is the shift of the HXI detector due to translation (Δr) and rotation (γ) of the EOB expressed in the SAT reference frame; and the vector from SAT origin to RAW origin in nominal condition is $\mathbf{r}_{0RAW} = \mathbf{r}_{HXI} - R_{ACT}^{SAT}\mathbf{r}_{RAW-HXI}^{ACT}$, where $\mathbf{r}_{HXI}$ is defined in Table 3 and Fig. 14; $\mathbf{r}_{RAW-HXI}^{ACT}$ is as per Fig. 15.

The coordinate rotation matrices are:

$$R_{RAW}^{ACT} = R_z(\gamma)$$
$$R_{SAT}^{ACT} = R_z(-\pi - \beta_H) \quad (10)$$
$$R_{ACT}^{SAT} = R_z(\pi + \beta_H)$$

The vectors used in this equation are displayed in Fig. 18.

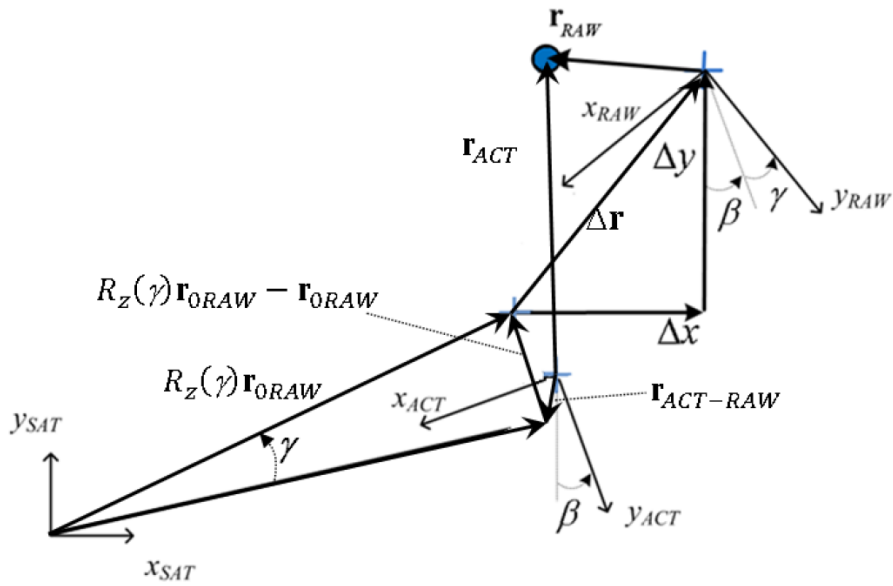

**Fig. 18** 2D-transformation from RAW to ACT coordinates. The blue circle represents the HXI measurement event.

When the data from both sensors are obtained in stable ACT coordinate frames, a rotation and superposition is used to generate the final image data obtained from both sensors in the SAT coordinate frame.

## 5  Flight Data Results

The time-binned data from the HXT observations of the Crab and G21.5-0.9 were processed using the algorithm described in Section 4. The lateral positions of binned, photon registration events, were first corrected for attitude pointing and then for deflection using CAMS data in their respective coordinate system.

Although the Crab and G21.5-0.9 are diffuse X-ray sources that extend at least several tens of arc seconds, the time binning effectively converts the different positions of photons arriving within the time bin period into a centroid location roughly coincident with the center of each X-ray source. After this binning, the data are comparable to a point source with equivalent brightness. It should be noted that the inherent blur of the HXT (several tens of arc seconds) is also reduced drastically for the time binned centroids as opposed to individual events. Although the binning reduces astronomical information about the source, the importance of the CAMS measurements and corrections could be validated.

In each time bin interval, there is a significant number of events (detected photons) landing on the detector. These photons are distributed across the detector because of intrinsic source extent (the source is not a point), telescope blur (due to limited point spread function of the optics), and motions due to spacecraft attitude change and EOB flexing. The time binning effectively averages the x/y positions of all photons arriving during the interval (to be more precise we adopt the median value). Therefore, the distribution of time binned points will typically be much smaller than the spread in the initial points. This holds if the spread is truly random, which is reasonable for the effects of telescope blurring and the extension of the X-ray source. However, the motions due to flexing of EOB and attitude control are effectively averaged only if they are occurring on timescales shorter than the binning interval. Thus, only relatively fast motions are averaged by the time binning process.

This also can be used to assess the validity of initial offset parameters $r_1^{CAMS1}$ and $r_2^{CAMS2}$ that were first obtained during the ground alignment. To estimate the nominal offset of the CAMS units with the goal of improving CAMS image correction several parameters are considered:

a. $r_{12}$ vector length (representing the distance between the two retro-reflectors, which is nominally 600 mm)

b. Rotation of the EOB as calculated using HXI gathered observations

c. Minimization of the spread of the time binned image on both HXIs equivalent to a point source.

The algorithm adopted for this minimization is the Affine Invariant Markov Chain Monte Carlo (MCMC) Ensemble sampling. The reason for this choice is the ability to constrain for physical parameters while leaving a trace of its progress in a form of a probabilistic distribution of the most likely answer. Fig. 19 shows the minimization for the combination of "b" and "c". Other attempts included minimizing "a", "b" and "c" independently, and minimizing their combinations. The EOB deployed average positions are estimated based on either HXI data for the two observations combined or from the CAMS measurements for the same observation. These are shown in Table 6.

Table 6 EOB average deployed position estimates where the zero positions correspond to the centers of the field of view for HXI detectors and CAMS sensors based on ground calibration.

| Estimate Source | X deviation (mm) | Y deviation (mm) | Rotation (mrad) |
|---|---|---|---|
| HXI Data | 0.504 | 0.732 | -2.61 |
| CAMS with Minimization for initial offsets | 0.513 | 0.759 | -2.62 |

Item "a" was also modified to minimize for optimal $r_{12}$ vector length and optimal distribution of HXI observations. The interesting finding about this minimization is that the solution returned a set of CAMS 1/2 linear trajectories that provided identical HXI image corrections. The result is the CAMS EOB correction quality is independent of its initial position estimate.

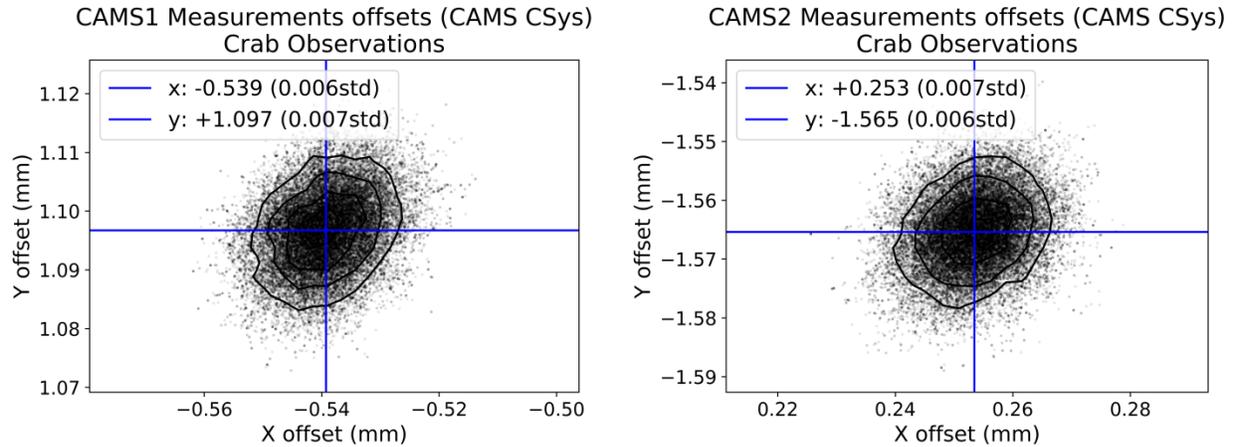

**Fig. 19** CAMS initial offset minimization.

The correction applied to the HXI time binned data is expected to reduce the data spread due to EOB deformations expressed in terms of standard deviation of binned photon coordinates relative to the image center. These standard deviation values for each observation are shown in Table 7. For each instrument and lateral direction, each row in Table 7 demonstrates the sequential improvement to the HXI observation correction. Some improvement comes from attitude correction, given the relatively large attitude swings (Fig. 20). Fig. 20 shows the variation of the raw centroid positions, attitude correction, and CAMS measurement during the same time frame for the Crab observation. The centroid of the distribution of Crab photons detected by each HXI in a 1-minute sampling period (blue crosses) is compared to the spacecraft attitude pointing deviation (green line) and CAMS measured optical structure deviation (red line). The readings are converted into detector pixel deviation as the desired outcome would be to have the correct photon centroid lie within 1-pixel from the center of the detector. Although, CAMS correction is more subtle than the attitude correction, it does reduce the spread in the image equivalent to a point source by removing EOB deflection component as reported in Table 7. Fig. 21 graphically represents the spread of the data points at different correction levels for observations of both targets. The HXI reported coordinate are converted to the spacecraft coordinate system to ease analysis (Fig. 21). For the Crab observation, one can see that the CAMS correction diminishes the spread in the data points making the final distribution smaller and more symmetric. As noted earlier, the data used in this analysis were aggregated using time binning periods of 60 seconds for the Crab and 300 seconds for G21.5-0.9. The total numbers of those periods within each observation are 136 for Crab and 396 for G21.5-0.9. These relatively large sampling periods are driven by the relatively small photon counts detected and to improve photon "centroid" calculation. Despite the larger time binning, the G21.5-0.9 observation contained much fewer data points per time bin, typically 400 counts compared to 20000 counts for the Crab observation, therefore, the centroiding was less efficient and the corrections are limited by residual diffusion from the source and the HXT blur. This explains the larger standard deviation and more diffuse time binned data points even after correction.

**Table 7** Standard deviations in x/y of the time-binned data points (centroids) during each observation and the impact of attitude and CAMS corrections on reducing the standard deviations.

|  |  | Crab (60sec) Standard deviation | | | G21.5-0.9 (300sec) Standard deviation | | |
|---|---|---|---|---|---|---|---|
|  |  | **Pixel** | **Arcsec** | **Improvement** | **Pixel** | **Arcsec** | **Improvement** |
| HXI1 X | Raw | 0.731 | 3.141 | - | 1.003 | 4.310 | - |
|  | Attitude | 0.621 | 2.668 | 15.0% | 1.069 | 4.593 | -6.6% |
|  | Att+CAMS | 0.372 | 1.598 | 40.1% | 1.010 | 4.340 | 5.5% |
| HXI2 X | Raw | 0.745 | 3.201 | - | 1.110 | 4.770 | - |
|  | Attitude | 0.596 | 2.561 | 20.0% | 1.121 | 4.817 | -1.0% |
|  | Att+CAMS | 0.388 | 1.667 | 34.9% | 1.001 | 4.301 | 10.7% |
| HXI1 Y | Raw | 0.658 | 2.827 | - | 2.257 | 9.698 | - |
|  | Attitude | 0.668 | 2.870 | -1.5% | 1.247 | 5.358 | 44.7% |
|  | Att+CAMS | 0.610 | 2.621 | 8.7% | 1.233 | 5.298 | 1.1% |
| HXI2 Y | Raw | 1.013 | 4.353 | - | 2.563 | 11.013 | - |
|  | Attitude | 0.366 | 1.573 | 63.9% | 1.130 | 4.856 | 55.9% |
|  | Att+CAMS | 0.304 | 1.306 | 16.9% | 1.087 | 4.671 | 3.8% |

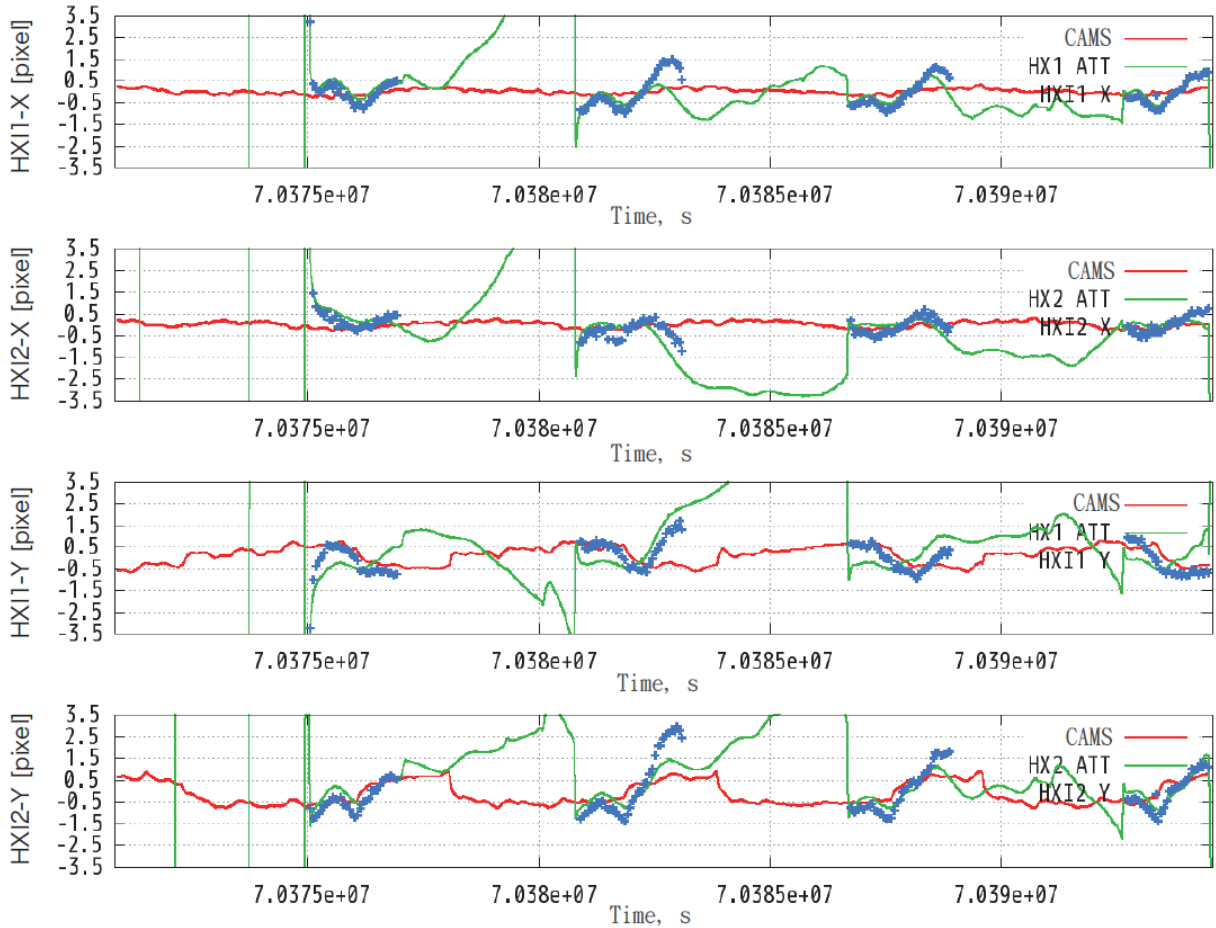

**Fig. 20** Raw data comparing the variations in the centroid position of the Crab in the HXI (blue), the spacecraft attitude (green) and the CAMS measurements (red). One pixel corresponds to 4.297 arcsec.

One would anticipate the improvement to be similar along the X and Y axes, however the CAMS improvement was found to be better in the X direction than in the Y. For example, in the Crab observation the improvement was ~40% in the X direction and ~10% in the Y (Table 7). This can be understood as arising from the telescope heater cycling observation discussed in Section 3.1 that was more substantial in the Y-direction. The temperature drifts in the CAMS instrument are less than 1.5 degrees for these two observations, therefore, the expected CAMS accuracy after temperature correction described earlier should be in the range of ~0.25 arcsec. The resulting value of about 2.6 arcsec for Crab thus contains top plate flexing contribution (in the range of 1.5 arcsec) along with other unaccounted factors.

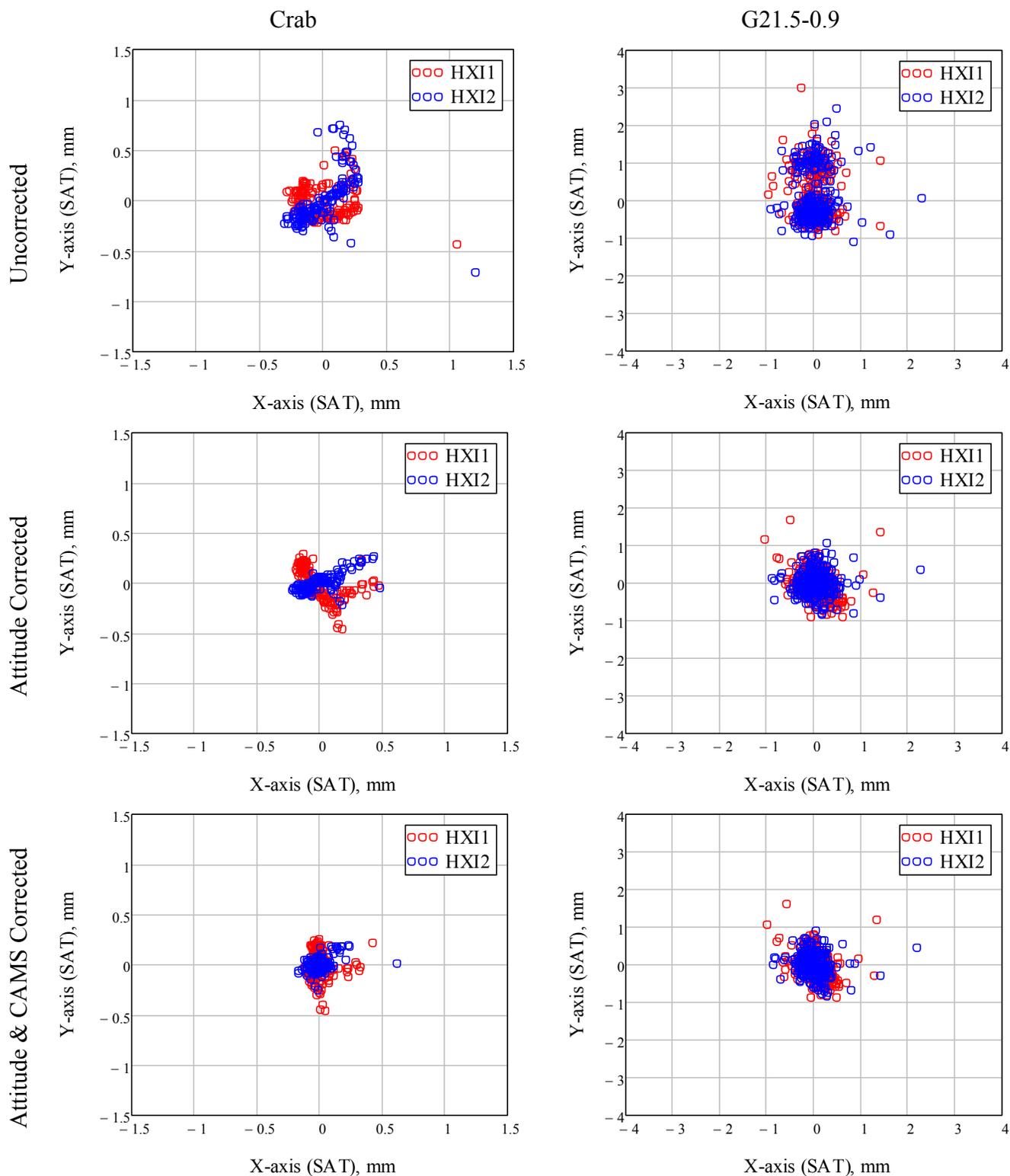

**Fig. 21.** Correction of the HXI observations of the Crab (left) and G21.5-0.9 (right). One mm shift corresponds to 17.19 arcsec angular shift in the focal plane (at HXI).

The CAMS performance can be assessed using these two observations. If the attitude and thermal effects experienced by the EOB structure occur on timescales shorter than the time binning intervals, applying a correction to individual photons would have likely resulted in a better overall improvement. However, to validate this assumption a bright point source should be used.

For extended sources like the Crab and G21.5-0.9, individual photons are dispersed 17-20 arcsec. The CAMS correction based on binned data is in the range of ~1 arcsec. The CAMS correction improves the initial image spread as the root-mean-square, hence its impact to the overall image size is negligible. Having either a true point source or an extended source with well-defined features would be essential to illustrate the CAMS correction using individual photon data.

For better illustration of image reconstruction, the binned HXI data to reconstruct equivalent point-source X-ray images (time binned images) by representing each binned data point with a Gaussian distribution with a specific width for notional image blur. These images are obtained using the following expression:

$$I(x,y) = \sum_{i=1}^{N} A_i \exp\left(\frac{-(X_i - X_0 - x)^2 - (Y_i - Y_0 - y)^2}{2\sigma^2}\right) \tag{11}$$

where $I(x,y)$ is the resulting reconstructed image; $N$ is the total number of binned data points; $A_i$ is the counts per data point; $X_i$ and $Y_i$ are median grouped coordinates for each data point; $X_0$ and $Y_0$ are the mean values of median grouped coordinates for the observation; and $\sigma$ is the assumed blur associated with each centroid data point. The reconstruction of these images would ideally have resulted in point sources limited by the blur and residual point spread after the time binning. The images are not expected to retain any structure or features of the actual Crab and G21.5-0.9 sources, as they are lost due to the time binning process.

The reconstructed time binned images are presented in Fig. 22. The adopted blur value was 0.05 mm or 0.86 arcsec. Centered images from both HXI detectors are added to generate them. One may notice that while attitude correction brings significant change in distribution of the data points, especially in its wings, the CAMS correction plays an important role for the shape of the observed source. The impact of the CAMS correction certainly depends on the blur. The impact is assessed by monitoring the full-width-half-maximum (FWHM) of the distribution (root mean square of $x$ and $y$ values) while changing the assumed blur of the telescope ($\sigma$) for the time binned Crab data (Fig. 23). The FWHM is preferred as an indicator rather than the standard deviation of data points used in Table 7, because it considers the different amount of counts per data point. The actual time binned images calculated for different values of assumed blur are shown in Fig. 24. This analysis demonstrates that while the attitude correction reaches its limitation of how much it improves at a level of 6 arcsec, the addition of CAMS continues improving the FWHM up to 2 arcsec. Therefore, the CAMS metrology becomes more important for higher resolution X-ray imaging.

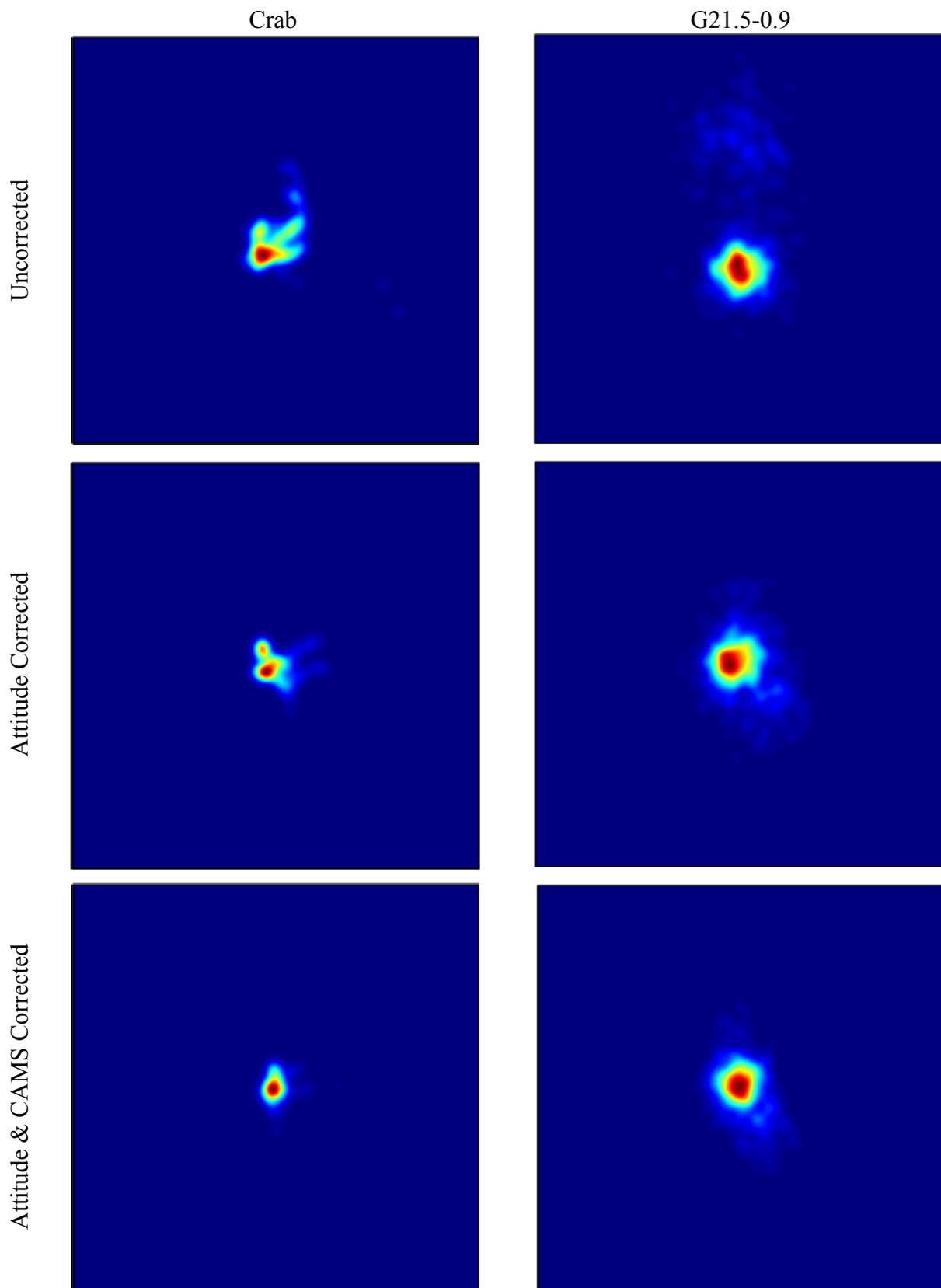

**Fig. 22** Distribution of time binned data for X-ray images of Crab and G21.5-0.9 observations based on averaging (binning) over 60 seconds (Crab) or 300 seconds (G21.5-0.9) at different levels of image correction. The image field of view is 68.75 arcsec. The assumed blur of each centroid data point is 0.86 arcsec.

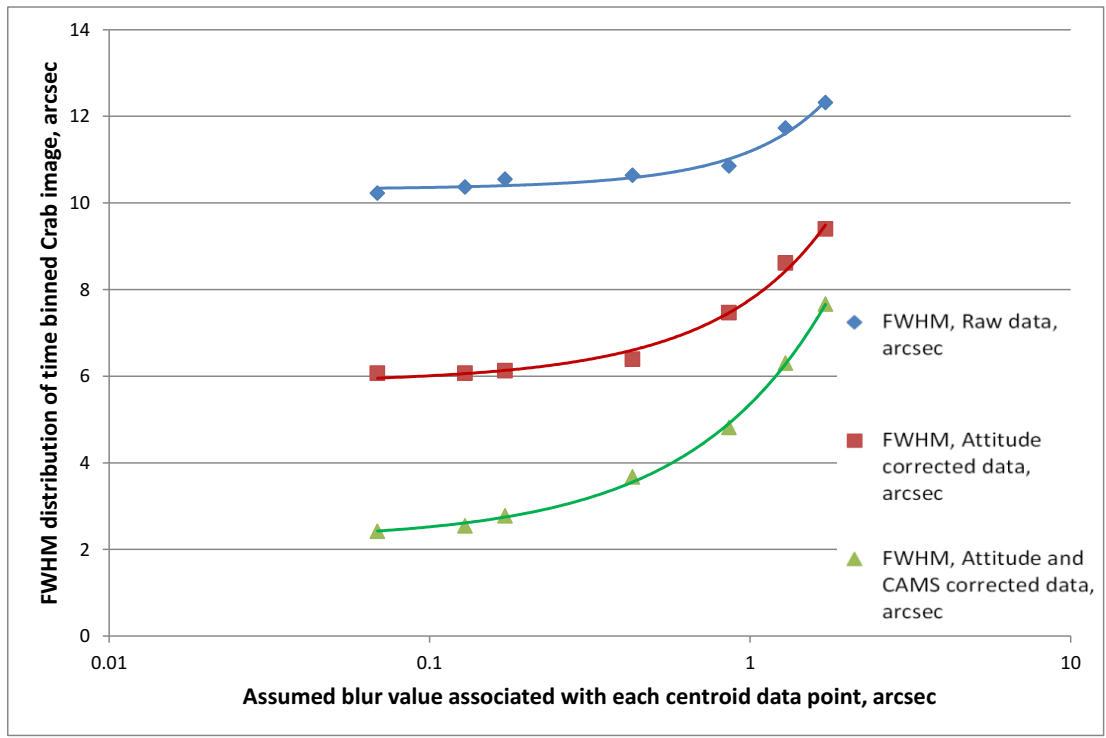

**Fig. 23** The resulting FWHM of the time binned Crab source distribution for raw (blue diamonds), attitude corrected (red squares) and attitude + CAMS corrected (green triangles) images as a function of the blur value used to generate the images.

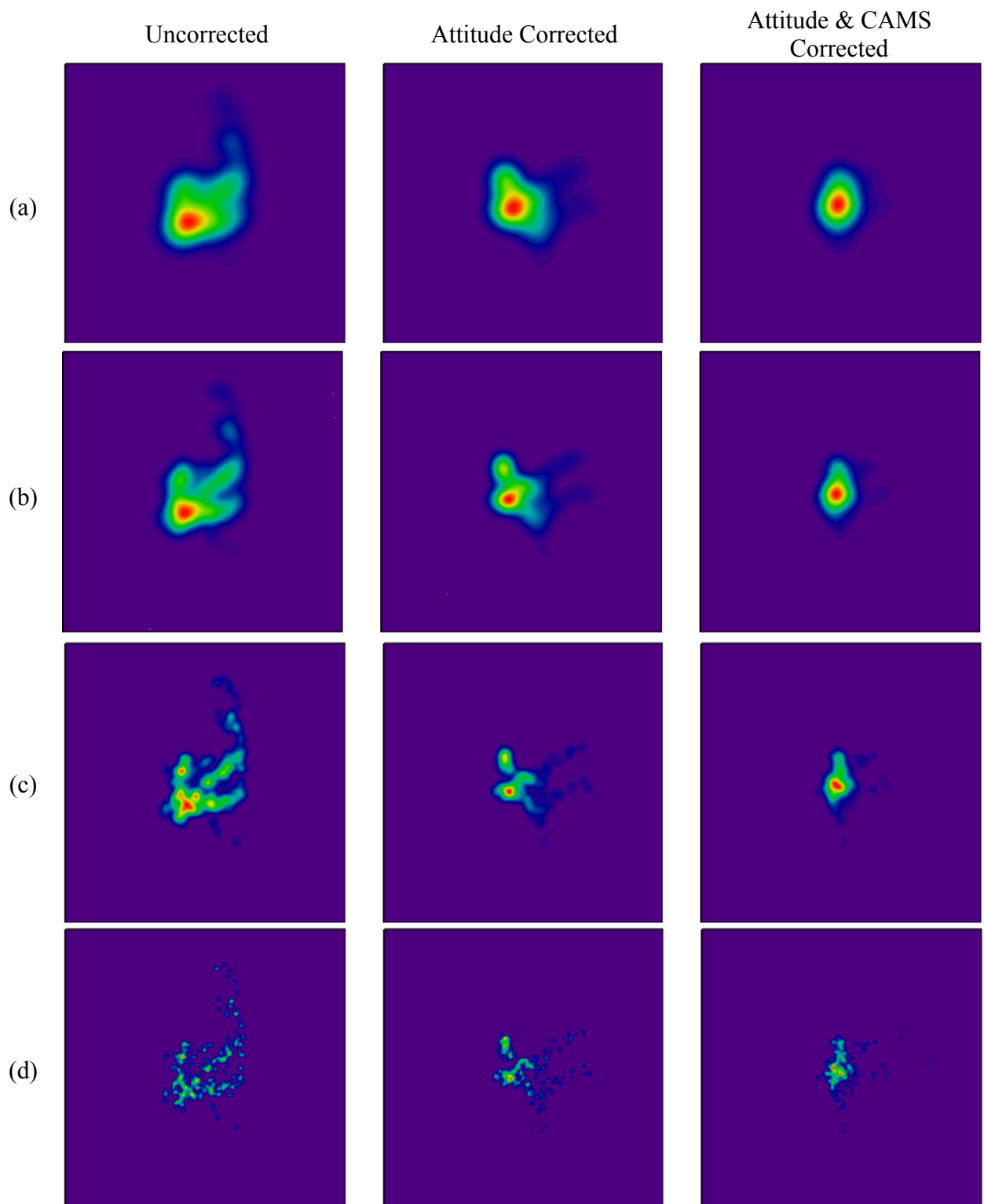

**Fig. 24** Time binned images of Crab observation with different correction levels and different assumptions on blur value associated with each centroid data point: (a) 1.29 arcsec; (b) 0.86 arcsec; (c) 0.43 arcsec; (d) 0.17 arcsec. The image field of view is 34.37 arcsec.

## 6 Conclusions

CAMS operations over the short duration of the Hitomi mission were successful, achieving micrometer resolution for the lateral shift in the EOB. In addition to its primary purpose the CAMS was employed in unexpected manners. The CAMS aided in the initial deployment of the extensible optical bench, providing crucial real-time information that was not otherwise available. It also provided highly precise information that could be used to understand the dynamics of the spacecraft structure. Such information was never acquired before.

The primary function of the CAMS was to improve HXI imaging quality by measuring displacement and rotation in the EOB deformation. Using astronomical observations of the Crab and G21.5-0.9, it is demonstrated that when the CAMS correction is applied in conjunction with the attitude correction, the HXI images are always improved. The CAMS measurement precision is limited by external factors such as the deformation of the FOB structure, where the CAMS-LD was mounted, due to the telescope optics heater cycling. It is also demonstrated that the CAMS correction becomes more important for higher resolution X-ray imaging.

The necessity to achieve long focal lengths while limiting costs by launching compact, light-weight structures, makes the demand for similar metrology systems widespread in space astronomy. A similar alignment system[8-10] is used on the NuSTAR satellite and it will continue to be required for several upcoming missions[11-13] in X-ray astronomy.

*Acknowledgments*

The authors are grateful to the entire CAMS team at the Canadian Space Agency and Neptec Design Group for their dedication to the project. We are also thankful to the JAXA team for their help in developing the CAMS. Thanks to Hans Krimm and Lorella Angelini for incorporating the CAMS correction into the pipeline. We are very appreciative to Prof. Tadayuki Takahashi for his patience and leadership. LCG acknowledges funding from the Canadian Space Agency.

*References*

1. T. Takahashi et al., "The Astro-H (Hitomi) x-ray astronomy satellite," in Space Telescopes and Instrumentation 2016: Ultraviolet to Gamma Ray, J. A. den Herder, T. Takahashi, M. Bautz, Ed., *Proc. SPIE* **9905**, (2016) [doi:10.1117/12.2232379].

2. H. Awaki et al., "Performance of the Astro-H hard x-ray telescope (HXT)," in Space Telescopes and Instrumentation 2016: Ultraviolet to Gamma Ray, J. A. den Herder, T. Takahashi, M. Bautz, Ed., *Proc. SPIE* **9905**, (2016) [doi:10.1117/12.2231258].

3. K. Nakazawa et al., "The hard x-ray imager (HXI) onboard Astro-H," in Space Telescopes and Instrumentation 2016: Ultraviolet to Gamma Ray, J. A. den Herder, T. Takahashi, M. Bautz, Ed., *Proc. SPIE* **9905**, (2016) [doi:10.1117/12.2231176].

4. L. Gallo et al., "The Canadian Astro-H metrology system," in Space Telescopes and Instrumentation 2012: Ultraviolet to Gamma Ray, J. A. den Herder, T. Takahashi, M. Bautz, Ed., *Proc. SPIE* **8443**, (2012) [doi:10.1117/12.926371].

5. L. Gallo et al., "The Canadian Astro-H metrology system," in Space Telescopes and Instrumentation 2014: Ultraviolet to Gamma Ray, J. A. den Herder, T. Takahashi, M. Bautz, Ed., *Proc. SPIE* **9144**, (2014) [doi:10.1117/12.2054921].

6. A. Koujelev, L. C. Gallo, and S. Gagnon, "Canadian Astro-H Metrology System," in *Optical Payloads for Space Missions*, S-E Qian, Ed., John Wiley & Sons, Ltd, Chichester, UK (2015) [doi:10.1002/9781118945179].

7. K. Ishimura et al., "Induced vibration of high-precision extensible optical bench during extension on orbit," *Japan Society for Aeronautical and Space Science (JSASS),* **submitted** (2017)

8. C. C. Liebe et al., "Metrology system for measuring mast motions on the NuSTAR mission," *Proc. Sensors Journal IEEE,* **12**(2006-2013), (2012)

9. C. C. Liebe et al., "Calibration and alignment of metrology system for the Nuclear Spectroscopic Telescope Array mission," *Opt. Eng.* **51**(4), 043605 (2012)

10. K. Forster et al., "Getting NuSTAR on target: predicting mast motion," in *Observatory Operations: Strategies, Processes, and System VI*, A. Peck, R. Seaman, C. Benn, Ed., *Proc. SPIE* **9910**, (2016) [doi:10.1117/12.2231239].

11. R. K. Smith et al., "Arcus: an ISS-attached high-resolution x-ray grating spectrometer," in Space Telescopes and Instrumentation 2014: Ultraviolet to Gamma Ray, J. A. den Herder, T. Takahashi, M. Bautz, Ed., *Proc. SPIE* **9144**, (2014) [doi:10.1117/12.2062671].

12. K. Mori et al., "A broadband x-ray imaging spectroscopy with high-angular resolution: the FORCE mission," in Space Telescopes and Instrumentation 2016: Ultraviolet to Gamma Ray, J. A. den Herder, T. Takahashi, M. Bautz, Ed., *Proc. SPIE* **9905**, (2016) [doi:10.1117/12.2231262].

13. X. Barcons et al., "Athena: ESA's X-ray observatory for the late 2020s," *Astron. Nach.* **338**(153), (2017)